\documentclass[pra,twocolumn,aps,showpacs,nofootinbib]{revtex4-1}

\usepackage{physics}
\usepackage{amsmath}
\usepackage{amssymb}
\usepackage{graphicx}
\usepackage{float}
\graphicspath{{PNG/}}
\usepackage{xcolor}
\usepackage{multirow}
\usepackage[utf8]{inputenc}

\makeatletter
\newcommand*\bigcdot{\mathpalette\bigcdot@{.5}}
\newcommand*\bigcdot@[2]{\mathbin{\vcenter{\hbox{\scalebox{#2}{$\m@th#1\bullet$}}}}}
\makeatother

\begin{document}
\title{Effects of Efimov states on quench dynamics in a three-boson trapped system}

\author{A. D. Kerin and A. M. Martin}
\affiliation{School of Physics, University of Melbourne, Parkville, VIC 3010, Australia}

\date{\today}

\begin{abstract}
We investigate the effects of Efimov states on the post-quench dynamics of a system of three identical bosons with contact interactions, in a spherically-symmetric three-dimensional harmonic trap. The quench we consider is in the s-wave contact interaction and we focus on quenches from the non-interacting to strongly interacting regimes and vice-versa. The calculations use the hyperspherical solutions of the three-body problem enable us to evaluate the semi-analytical results of the Ramsey and particle separation, post quench. In the case where the interactions are quenched from the non-interacting to strongly interacting regime we find convergent aperiodic solutions for both the Ramsey signal and the particle separation. In contrast for quenches from the strongly interacting regime to the non interacting regime both the Ramsey signal and particle separation are periodic functions. However, in this case we find that the solutions for the particle separation diverges indicating that in such a system large oscillations may be observable.

\end{abstract}
\maketitle

\section{Introduction}
\label{sec:Intro}

Efimov states are a unique type of many-body quantum state where short-range interactions create effective long-range forces due to the exchange interaction. They were first predicted by Efimov \cite{efimov1971bound} and first observed by Kraemer {\it et al.} \cite{kraemer2006evidence}. Efimov states appear in systems of as few as three bodies \citep{kartavtsev2007low, kartavtsev2008universal, endo2011universal, petrov2012few,werner2008trapped, jonsell2002universal,kerin2022energetics} and while the circumstances under which Efimov states appear is well understood the specifics of those states, e.g. their energies, can vary depending on the specifics of the system. The influence of Efimov states is highly relevant to a number of topics of cold-gas research including their effects on, among other quantities, the two and three-body contacts or three-body decay rates after a quench \cite{fletcher2017two, colussi2018dynamics, eigen2018universal, d2018efimov, colussi2019bunching, musolino2019pair, musolino2022bose}.

In this work we consider the dynamics of a system of three identical bosons, interacting via a contact interaction, in an isotropic harmonic trap. Such systems can be constructed in experiment \cite{serwane2011deterministic, murmann2015two, zurn2013pairing, zurn2012fermionization, PhysRevLett.96.030401} in the form of dilute ultracold gases. Specifically, we consider the dynamics of the system after the a quench in the contact s-wave interactions. We consider two quench pathways, from the non-interacting regime to the strongly interacting (unitary) regime and vice-versa. We utilise known solutions of the static case \cite{Cui2012,d2018efimov, jonsell2002universal, blume2010breakdown, kerin2022energetics, kestner2007level} to calculate the Ramsey signal and particle separation as functions of time following the quench and investigate the effects of different Efimov energy spectra upon the dynamics. Such solutions have been used to calculate thermodynamics quantities such as virial coefficients or Tan contacts \cite{PhysRevLett.102.160401, liu2010three, PhysRevLett.96.030401, Cui2012, PhysRevA.85.033634, PhysRevLett.107.030601, mulkerin2012universality, PhysRevA.86.053631, Nature463_2010, Science335_2010, levinsen2017universality, daily2010energy, bougas2021few, colussi2019two, colussi2019bunching, enss2022complex}. In this work we consider a three-dimensional system complementing previous investigations into quench dynamics in two-dimensional \cite{bougas2022dynamical} and one-dimensional \cite{pecak2016two,volosniev2017strongly,kehrberger2018quantum, sowinski2019one} systems.	

We note that the predictions in this paper are experimentally testable with current techniques. Notably Ref. \cite{guan2019density} prepared a harmonically trapped system of \textit{two} {}\textsuperscript{6}Li atoms, quenched in \textit{trap geometry} and measured the particle separation. It is also possible to experimentally obtain systems of three harmonically trapped atoms \cite{serwane2011deterministic, murmann2015two, zurn2013pairing, zurn2012fermionization, PhysRevLett.96.030401} and the quench in s-wave scattering length is possible using tools such as Feshbach resonance \cite{fano1935feshbackh, feshbach1958feshbackh, tiesinga1993feshbackh, chin2010feshbach}. Additionally experiments measuring the Ramsey signal of trapped cold gases after a quench have been performed \cite{cetina2016ultrafast}.

This paper is structured in the following way. Sec. \ref{sec:Overview} provides an overview of the hyperspherical solution to the problem of three identical bosons in a spherical harmonic trap interacting via a contact interaction, including a review of Efimov states. In Sec. \ref{sec:Quench} we use the static solutions to calculate observables of the post-quench system. We consider the non-interacting to unitary (forwards) and vice versa (backwards) quenches. In these two cases we calculate the Ramsey signal, the overlap of the pre- and post-quench states, and the expectation of the particle separation. For the forwards quench we find that both quantities can be calculated semi-analytically and in the reverse case the Ramsey signal is still calculable but the particle separation diverges.

\section{Overview of the Three-Body Problem}
\label{sec:Overview}
To begin, the Hamiltonian of three identical non-interacting bodies in an isotropic three-dimensional harmonic trap is
\begin{eqnarray}
\hat{H}=\sum_{k=1}^{3} \left[ \frac{-\hbar^2}{2m}\nabla_{k}^2 +\frac{m \omega^2 r_{k}^2}{2}\right],
\end{eqnarray}
where $\vec{r}_{k}$ is the position of the $k^{\rm th}$ particle, $m$ is the particle mass, and $\omega$ is the trapping frequency. For convenience we define the length scales
\begin{eqnarray}
a_{\mu}=\sqrt{\frac{\hbar}{\mu\omega}}, \quad a_{M}=\sqrt{\frac{\hbar}{M\omega}},
\end{eqnarray} 
where $\mu=m/2$, and $M=3m$.

We use the Bethe-Peierls boundary condition to model the contact interactions \cite{bethe1935quantum}
\begin{eqnarray}
\lim_{r_{ij}\rightarrow0}\left[\frac{d(r_{ij}\Psi)}{dr_{ij}} \frac{1}{r_{ij}\Psi}\right]_{r_{ij}\rightarrow0}=\frac{-1}{a_{\rm s}},
\label{eq:BethePeierls}
\end{eqnarray}
where $\Psi$ is the total three-body wavefunction, $r_{ij}=|\vec{r}_{i}-\vec{r}_{j}|$, and $a_{\rm s}$ is the s-wave scattering length. 

The wavefunction of three identical harmonically trapped atoms subject to Eq. (\ref{eq:BethePeierls}) is known \cite{werner2006unitary, PhysRevLett.102.160401}. In particular the hyperspherical formulation \cite{werner2006unitary} is a closed form description of the wavefunction in the strongly interacting (unitary) and non-interacting regimes.

We define the hyperradius $R$ and hyperangle $\alpha$ 
\begin{eqnarray}
R^2=\sqrt{r^2+\rho^2},\quad \alpha=\arctan{(r/\rho)},
\end{eqnarray}
where 
\begin{eqnarray}
\vec{r}&=&\vec{r}_{2}-\vec{r}_{1}, \\
\rho &=& \frac{2}{\sqrt{3}}(\vec{r}_{3}-\frac{\vec{r}_{1}+\vec{r}_{2}}{2}),
\end{eqnarray}
and
\begin{eqnarray}
\vec{C}&=&\frac{\vec{r}_{1}+\vec{r}_{2}+\vec{r}_{3}}{3}
\end{eqnarray}
is the centre-of-mass (COM) coordiante. The COM Hamiltonian is a simple harmonic oscillator (SHO) Hamiltonian. The COM wavefunction is unaffected by Eq. (\ref{eq:BethePeierls}) and is a SHO wavefunction of argument $\vec{C}$ and lengthscale $a_{M}$. In hyperspherical coordinates the relative Hamiltonian is given
\begin{eqnarray}
\hat{H}_{\rm rel}&=&\frac{-\hbar^2}{2\mu}\Bigg( \frac{\partial^2}{\partial R} +\frac{1}{R^2\sin(\alpha)\cos(\alpha)}\frac{\partial^2}{\partial \alpha^2}\left(\cos(\alpha)\sin(\alpha)\right)\nonumber\\
&+&\frac{5}{R}\frac{\partial}{\partial R}-\frac{4}{R^2}-\frac{\hat{\Lambda}_{r}^2}{R^2\sin(\alpha)}-\frac{\hat{\Lambda}_{\rho}^2}{R^2\cos(\alpha)} \Bigg)+\frac{\mu\omega^2R^2}{2}.\nonumber\\\label{eq:HypersphericalHamiltonian}
\end{eqnarray}

We define an ansatz wavefunction of the form
\begin{eqnarray}
\psi_{\rm 3brel}&=&N_{qls}\frac{F_{qs}(R)}{R^2}(1+\hat{P}_{13}+\hat{P}_{23})\frac{\varphi_{ls}(\alpha)}{\sin(2\alpha)}Y_{lm}(\hat{\rho}),\label{eq:HypersphericalWavefunction}\qquad
\end{eqnarray}
where $N_{qls}$ is the normalisation constant, $F_{qs}$ is the hyperradial wavefunction and $\phi_{ls}=(1+\hat{P}_{13}+\hat{P}_{23})\varphi_{ls}(\alpha)Y_{lm}(\hat{\rho})/\sin(2\alpha)$ is the hyperangular wavefunction. The exchange operators $\hat{P}_{13}$ and $\hat{P}_{23}$ exchange the positions of particles one and three and particles two and three respectively. 

Requiring the ansatz to be an eigenfunction of the Hamiltonian leads to the hyperangular and hyperradial equations
\begin{eqnarray}
s^2\varphi_{ls}(\alpha)&=&-\varphi_{ls}''(\alpha)+\frac{l(l+1)}{\cos^2(\alpha)}\varphi_{ls}(\alpha),\label{eq:FirstCondition}\\
E_{\rm rel}F_{qs}(R)&=&\frac{-\hbar^2}{4\mu}\left(F_{qs}''(R)+\frac{F_{qs}'(R)}{R}\right)\nonumber\\
&+&\left(\frac{\hbar^2s^2}{4\mu R^2}+\mu\omega^2 R^2\right)F_{qs}(R),\label{eq:SecondCondition}
\end{eqnarray}
and noting that a divergence at $\alpha=\pi/2$ is non-physical gives the condition
\begin{eqnarray}
\varphi_{ls}\left(\frac{\pi}{2}\right)&=&0.\label{eq:ThirdCondition}
\end{eqnarray}

Eqs. (\ref{eq:FirstCondition})-(\ref{eq:ThirdCondition}) determine the form of $F_{qs}(R)$ and $\varphi_{ls}(\alpha)$ \cite{werner2006unitary, PhysRevA.74.053604, liu2010three}
\begin{eqnarray}
F_{qs}(R)&=&\begin{cases}
\left(\tilde{R}\right)^s e^{-\tilde{R}^2/2}L_{q}^{s}\left(\tilde{R}^2\right), & s^2>0\\
 & \\
\dfrac{1}{\tilde{R}} W_{\dfrac{E_{\rm rel}}{2\hbar\omega},\dfrac{s}{2}} (\tilde{R}^2), & s^2<0
\end{cases},
\\
\varphi_{ls}(\alpha)&=&\cos^{l+1}(\alpha)\nonumber\\
&\times&{}_{2}F_{1}\left(\frac{l+1-s}{2},\frac{l+1+s}{2};l+\frac{3}{2};\cos^2(\alpha)\right),\nonumber\\
\end{eqnarray}
where $L_{q}^{s}$ is the associated Laguerre polynomial, $W_{E_{\rm rel}/2\hbar\omega,s/2}$ is the Whittaker function, ${}_{2}F_{1}$ is the hypergeometric function, $\tilde{R}=R/a_{\mu}$, $q\in\mathbb{Z}_{\geq0}$, and $s^2\in\mathbb{R}$ are the energy quantum numbers and $l\in\mathbb{Z}_{\geq0}$ is the angular momentum quantum number. In the rest of this work we consider only the $l=0$ case for reasons that are elucidated in the appendix. As such we omit angular momentum indices ($l$) in subsequent notation.

In this framework the $s$-eigenvalues can only be fully specified in the non-interacting and unitary regime. In the unitary limit applying Eq. (\ref{eq:BethePeierls}) to Eq. (\ref{eq:HypersphericalWavefunction}) gives the transcendental equation
\begin{eqnarray}
0=\frac{d \varphi_{s}'(\alpha)}{d\alpha}\Big|_{\alpha=0}+\frac{8}{\sqrt{3}}\varphi_{s}\left(\frac{\pi}{3}\right),
\label{eq:Transcendental}
\end{eqnarray}
which determines the $s$-eigenvalues, some solutions are presented below in Table \ref{tab:sEigenvalues}. In the non-interacting limit applying Eq. (\ref{eq:BethePeierls}) to Eq. (\ref{eq:HypersphericalWavefunction}) gives $s$ as
\begin{eqnarray}
s=\begin{cases}
2 \\
2n+6 \\
\end{cases}
,
\end{eqnarray}
for $l=0$, where $n\in\mathbb{Z}_{\geq0}$.
\begin{table}[H]
\center
\begin{tabular}{|c|c|}

\hline
 $n$ & $s$ \\
\hline
0 & $i\cdot$ 1.006\dots \\
1 & 4.465\dots \\
2 & 6.818\dots \\
3 & 9.324\dots \\
\hline
\end{tabular}
\caption{The three-body $s$-eigenvalues at unitarity for the 3 boson case for $l=0$ to three decimal places.}
\label{tab:sEigenvalues}
\end{table}

\begin{figure}[H]
\includegraphics[height=5.5cm,width=8.5cm]{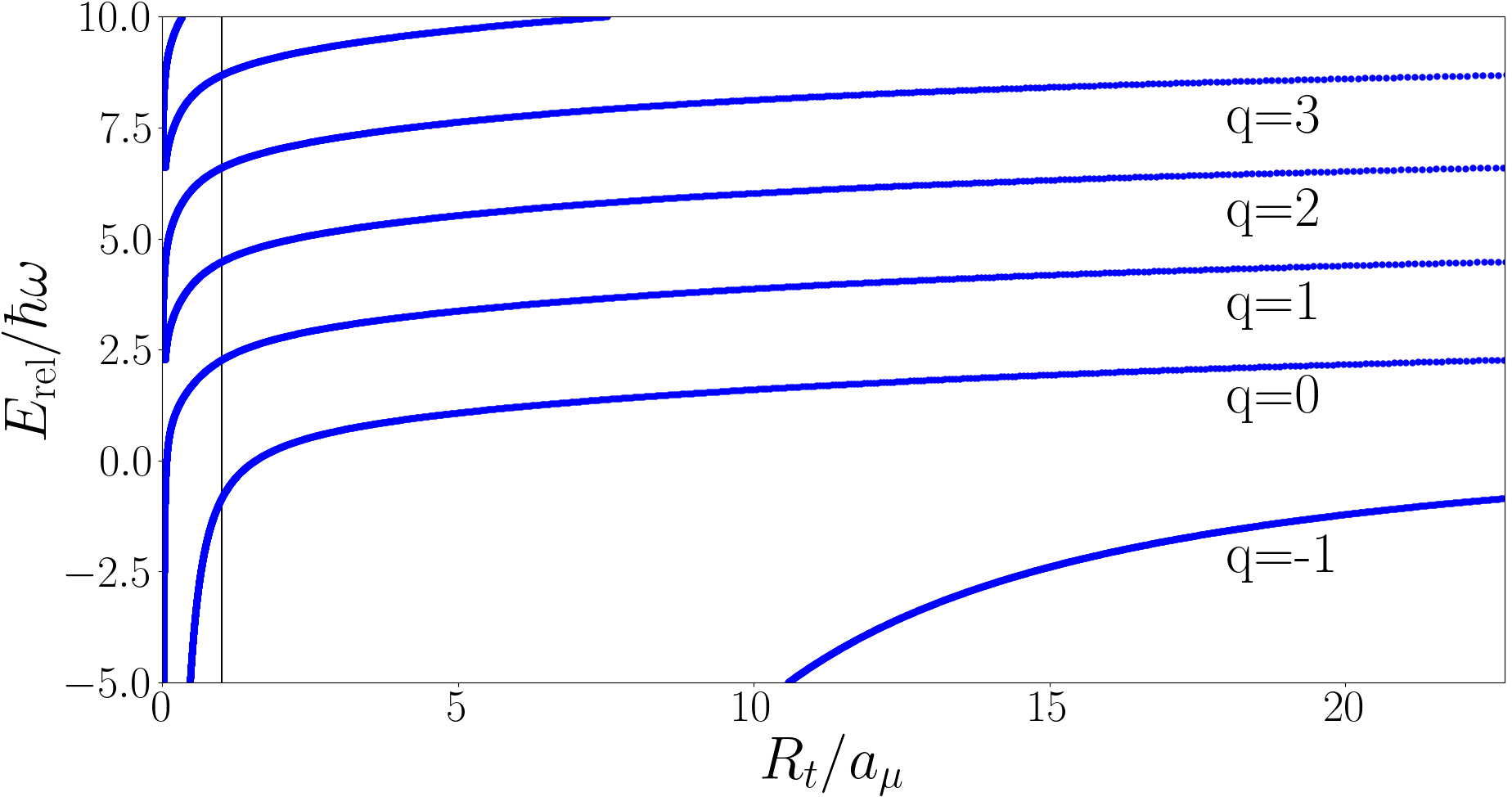}
\caption{The energy spectrum for Efimov states as defined by Eq. (\ref{eq:EfimovEnergies}). Calculated using $s=i\cdot 1.006\dots$. The upper limit on the horizontal axis is $R_{t}/a_{\mu}=e^{\pi/|s|}\approx22.7$, and the vertical black line is $R_{t}=a_{\mu}$. }
\label{fig:EfimovEspec}
\end{figure}

For $s^2>0$, the universal case, the energy of the wavefunction is $E_{\rm rel}=(2q+s+1)\hbar\omega$, which is implicitly determined by Eq. (\ref{eq:SecondCondition}), recall $q\in\mathbb{Z}_{\geq0}$. For $s^2<0$, the Efimov case, the energy is not uniquely determined by requiring the wavefunction, Eq. (\ref{eq:HypersphericalWavefunction}), be an eigenfunction of the Hamiltonian, Eq. (\ref{eq:HypersphericalHamiltonian}), it is instead a free parameter. Hence we require an additional condition to fix the energy. The Efimov hyperradial wavefunction oscillates increasingly rapidly as $R\rightarrow0$ and we set a condition to fix the phase of the oscillation \cite{werner2006unitary,jonsell2002universal}
\begin{eqnarray}
\arg \Gamma &&\left[\frac{1+s-E_{\rm rel}/\hbar\omega}{2} \right]\nonumber\\
&&=-|s|\ln(\frac{R_{t}}{a_{\mu}})+\arg \Gamma(1+s) \mod \pi, \label{eq:EfimovEnergies}
\end{eqnarray}
where $R_{t}$ is the three-body parameter, an arbitrary parameter with units of distance. $R_t$ determines the energies of the Efimov states. Physically speaking $R_{t}$ is required because, in the Efimov case, Eq. (\ref{eq:SecondCondition}) has an attractive potential term proportional to $1/R^2$ which allows for arbitrarily small interparticle distances. At small distances the contact interaction assumption breaks down and the short range nature of the interaction become significant. The Efimov energies are plotted as a function of $R_{t}$ in Fig. \ref{fig:EfimovEspec}. The energy spectrum is unbounded from below and above, we label the states with $q\in\mathbb{Z}$, defining the $q=0$ state to be the lowest energy state with $E_{\rm rel}>0$ at $R_{t}=\exp(\pi/|s|)a_{\mu}$. Note that the energy of the $q=N$ state evaluated at $R_{t}=a_{\mu}$ is equal to the energy of the $q=N-1$ state evaluated at $R_{t}=\exp(\pi/|s|)a_{\mu}$.

\section{Quench Dynamics}
\label{sec:Quench}
In this paper we calculate the Ramsey signal and particle separation after a quench in $a_{s}$. As part of this we need to calculate various integrals of the hyperspherical wavefunction and the details of these are presented in the appendix. The COM wavefunction is unaffected by Eq. (\ref{eq:BethePeierls}) and so is unaffected by a quench in $a_{\rm s}$. As such only the relative motion impacts the system behaviour.

The time-dependent post-quench relative wavefunction is given
\begin{eqnarray}
\ket{\psi(t)}&=&e^{-i\hat{H}_{\rm rel}t/\hbar}\ket{F_{q_{\rm i}s_{\rm i}}\phi_{s_{\rm i}}}\nonumber\\
&=&\sum_{q,s}
\bra{F_{qs}\phi_{s}}\ket{F_{q_{\rm i}s_{\rm i}}\phi_{s_{\rm i}}}e^{-iE_{qs}t/\hbar}\ket{F_{qs}\phi_{s}}, \quad
\end{eqnarray}
where $\hat{H}_{\rm rel}$ is the post-quench relative Hamiltonian, quantum numbers with subscript i refer to the initial state, quantum numbers with no subscripts are the post-quench eigenvalues and $E_{qs}$ are the post-quench eigenenergies.

\subsection{Ramsey signal}
\label{sec:RamseySignal}
The Ramsey signal \cite{cetina2016ultrafast} is defined as the wavefunction overlap of the initial and final states, 
\begin{eqnarray}
S(t)&=&\bra{\Psi_{\rm i}(t)}\ket{\Psi'(t)}=\sum_{j=0}^{\infty}|\bra{\Psi_{\rm i}(0)}\ket{\Psi'_{j}}|^2e^{-i(E'_{j}-E_{\rm i})t/\hbar},\qquad\label{eq:RamseyDefn}
\end{eqnarray}
where $\Psi_{\rm i}$ is the initial state with energy $E_{\rm i}$, $\Psi'$ is the post-quench state and the $\Psi'_{j}$s are the eigenstates of the post-quench system with energy $E'_{j}$, $j$ is summing over all post-quench eigenstates.

As mentioned above the COM wavefunction is unaffected by the quench and integrates to one. Hence the Ramsey signal depends only on the relative wavefunction,
\begin{eqnarray}
S(t)&=&\sum_{q,s}|\bra{F_{q_{\rm i}s_{\rm i}}\phi_{s_{\rm i}}}\ket{F_{qs}\phi_{s}}|^2e^{-i(E_{qs}-E_{q_{\rm i}s_{\rm i}})t/\hbar}.\label{eq:RamseySignal}\quad
\end{eqnarray}
To evaluate the Ramsey signal we need to evaluate the hyperradial integral $\bra{F_{qs}}\ket{F_{q_{\rm i}s_{\rm i}}}$ and the hyperangular integral $\bra{\phi_{s}}\ket{\phi_{s_{\rm i}}}$. The appendix contains the details of the evaluation of these integrals. With the integrals known we can then calculate the Ramsey signal for the forwards and backwards quenches. There is a degree of freedom in the choice of the Efimov energy spectrum, determined by the value of $R_{t}$. Whatever the value of $R_t$ the normalisation is preserved but the post-quench behaviour is nonetheless affected by the choice of $R_{t}$.

\begin{figure}[H]
\includegraphics[height=5.5cm, width=8.5cm]{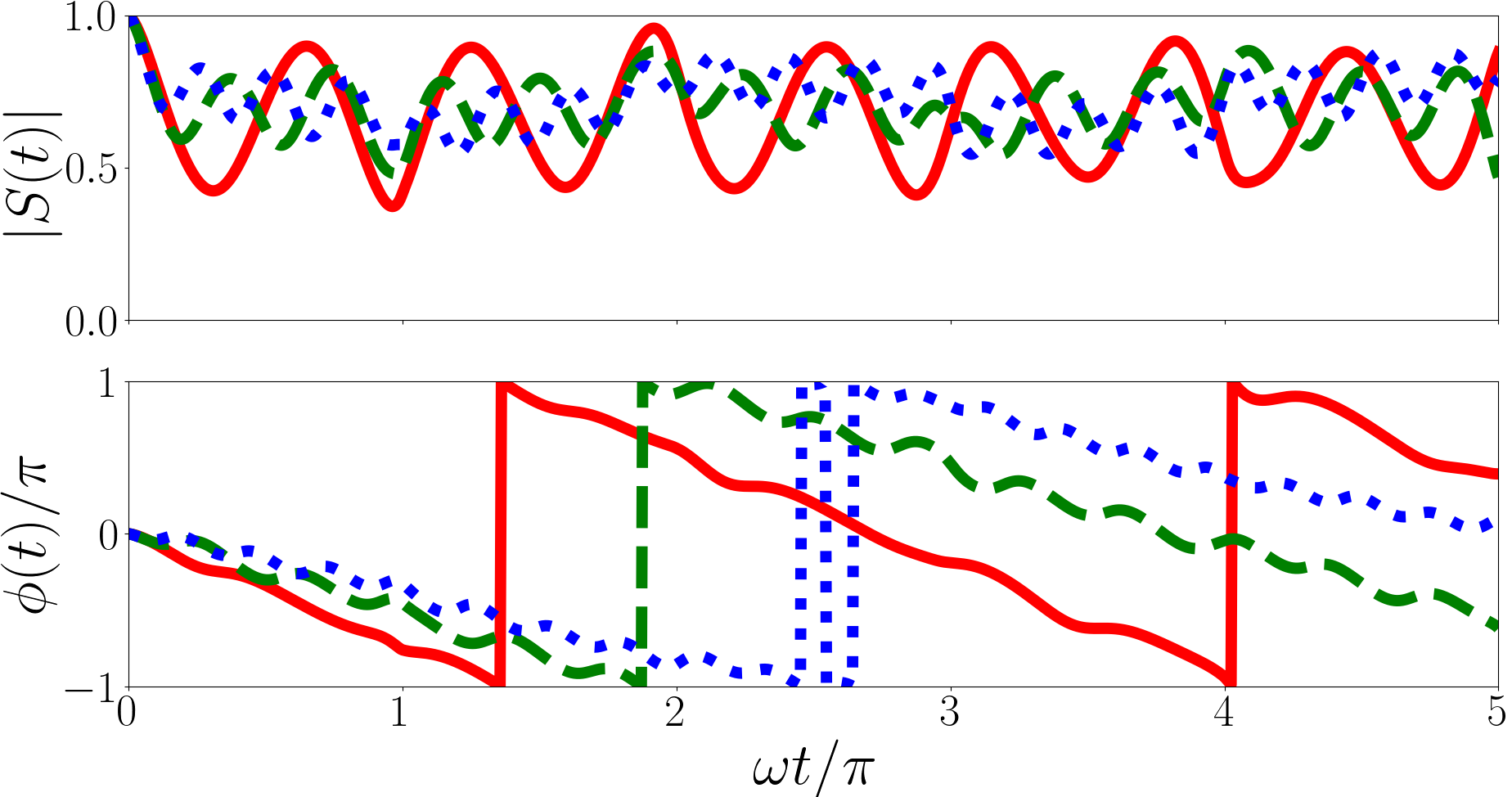}
\includegraphics[height=5.5cm, width=8.5cm]{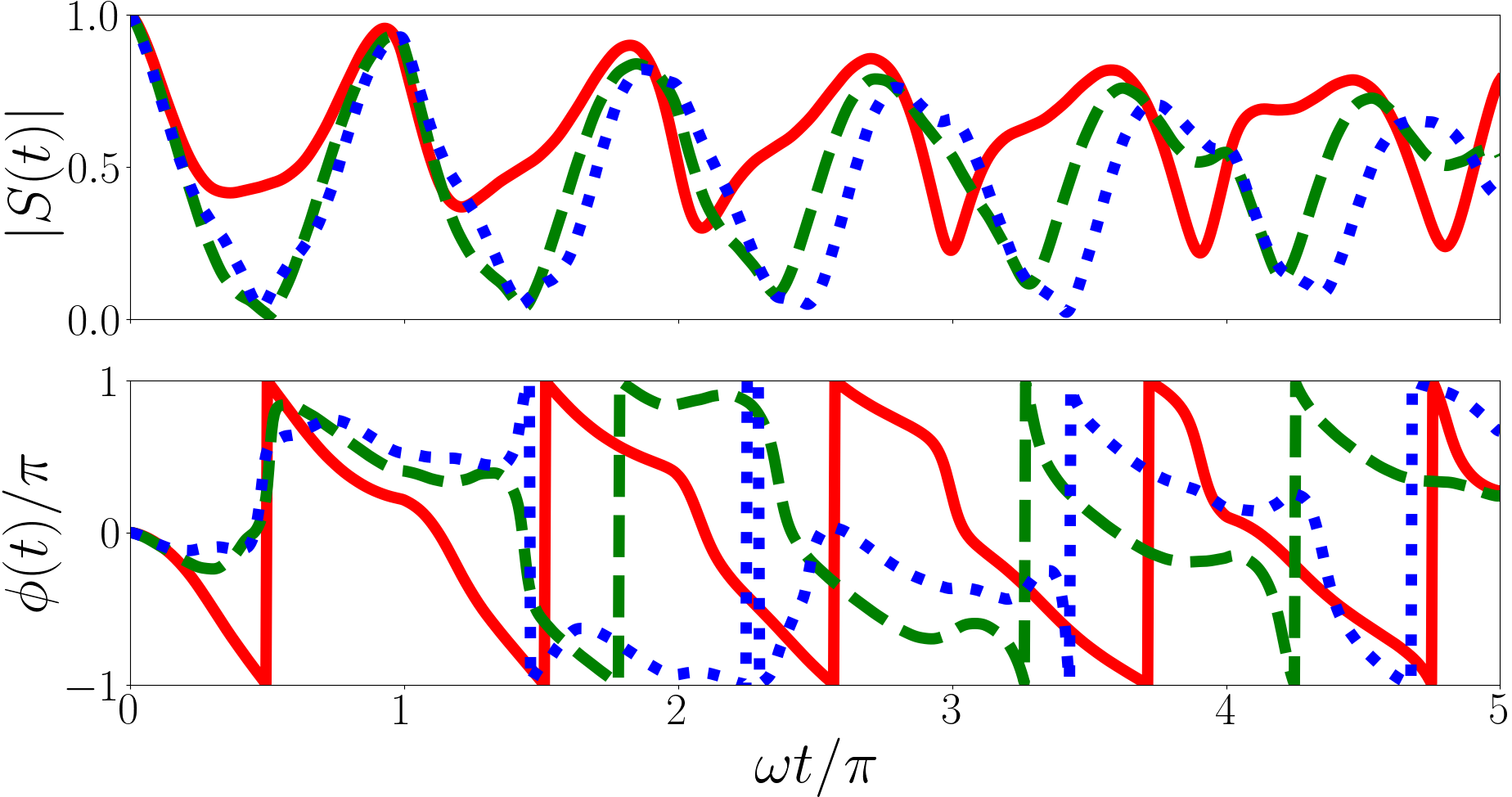}
\includegraphics[height=5.5cm, width=8.5cm]{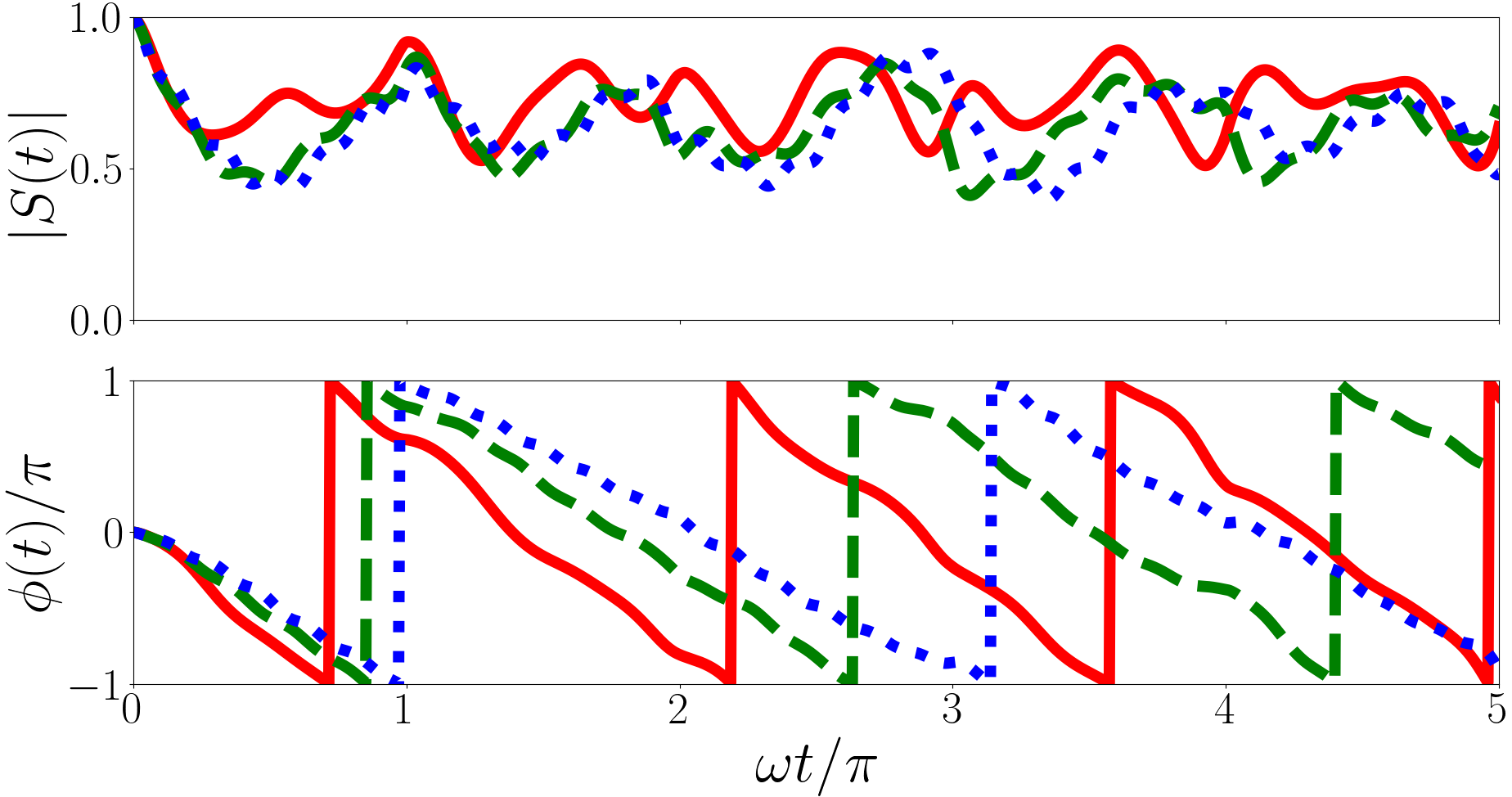}
\caption{Ramsey signal of the system quenched from non-interacting to unitarity. In each panel $s_{\rm i}=2$ and the solid red line corresponds to $q_{i}=0$, the dashed green line to $q_{i}=1$, and the dotted blue line to $q_{i}=2$. The upper panel uses $R_t=a_{\mu}$ to calculate the Efimov energy spectrum, the middle panel uses $R_t=5a_{\mu}$ and the lower $R_t=10a_{\mu}$. These Ramsey signals are evaluated using Eq. (\ref{eq:RamseyDefn}) with 40 terms in each of the sums, 1600 terms total. We find that the summation is convergent.}
\label{fig:RamseyForwards}
\end{figure}

The Ramsey signal is the weighted sum of oscillators, $S(t)=Ae^{-iat}+Be^{-ibt}+Ce^{-ict}+\dots$, where the weights are the square overlaps between initial state and post-quench eigenstates and the angular frequencies are the differences between the initial energy and post-quench eigenenergies. The magnitude is similarly a weighted sum of oscillators but the angular frequencies of the oscillatory terms are the differences between post-quench eigenenergies, $(a-b), (b-c), (a-c),\dots$. The \textit{phase} of the Ramsey signal is dominated by the phase of the most heavily weighted terms.

In Fig. \ref{fig:RamseyForwards} we plot the Ramsey signal of the forwards quench for a number of initial states and values of $R_{t}$. The calculations of the Ramsey signal for the forwards quench are performed including only the $q\geq-1$ Efimov energies except for the $R_{t}=a_{\mu}$ calculation which includes only the $q\geq0$ Efimov energies. The neglected energies are significantly lower (e.g. for $R_{t}=a_{\mu}$ the $q=-1$ Efimov energy is $\approx-566\hbar\omega$) and do not contribute meaningfully. Unlike with the two-body case \cite{kerin2020two} the magnitude of the Ramsey signal of the forwards quench is aperiodic. This is because the post-quench eigenenergies are irrational because the unitary $s$-eigenvalues are irrational as are the Efimov energies in general. This means the angular frequencies in Eq. (\ref{eq:RamseySignal}) ($a$, $b$, $c\dots$ from the previous paragraph) are irrational as are the differences between them, hence the magnitude and phase of the Ramsey signal are aperiodic. 

In the $R_{t}=a_{\mu}$, $(q_{\rm i},s_{\rm i})=(0,2)$ case (solid red line in the upper panel of Fig. \ref{fig:RamseyForwards}) the post quench states with the largest overlaps are $(E_{q=1},s)\approx (2.27\hbar\omega,i \cdot 1.006)$, with square overlap of $\approx 0.666$, $(E_{q=0}, s) \approx (-0.85\hbar\omega,i \cdot 1.006)$, with square overlap of $\approx 0.14$ and $(q,s)=(0,4.465\dots)$, with square overlap $\approx 0.105$. The two largest modes in the magnitude have periods of $\approx2\pi/3\omega$. The phase is dominated by a period of $\approx2.7\pi/\omega$.

In the $R_{t}=5a_{\mu}$, $(q_{\rm i},s_{\rm i})=(0,2)$ case (solid red line in the middle panel of Fig. \ref{fig:RamseyForwards}) the most significant terms are $(E_{q=0},s)\approx (1.077\hbar\omega,i\cdot 1.006)$,  with a square overlap of $\approx0.583$, $(E_{q=1},s)\approx(3.37\hbar\omega,i\cdot 1.006)$ with a square overlap of $\approx 0.24$ and $(q,s)=(0,4.465\dots)$ with a square overlap of $\approx 0.105$. The two largest modes in the magnitude have periods of $\approx0.87\pi/\omega$ and $\approx0.45\pi/\omega$. The phase is dominated by the period $\approx\pi/\omega$.

In the $R_{t}=10a_{\mu}$, $(q_{\rm i},s_{\rm i})=(0,2)$ case (solid red line in the lower panel of Fig. \ref{fig:RamseyForwards}) the terms with the largest overlaps are $(E_{q=0},s)\approx(1.603\hbar\omega,i\cdot 1.006)$ with a square overlap of $\approx 0.72$, $(q,s)=(0,4.465\dots)$ with a square overlap of $\approx 0.105$ and $(E_{q=1},s)\approx (3.875\hbar\omega,i\cdot 1.006)$ with a square overlap of $\approx0.092$. This leads to two main modes in the magnitude with periods of $\approx0.5\pi/\omega$ and $\approx0.9\pi/\omega$ and the phase is dominated by the period of $\approx1.4\pi/\omega$.

\begin{figure}[H]
\includegraphics[height=5.5cm, width=8.5cm]{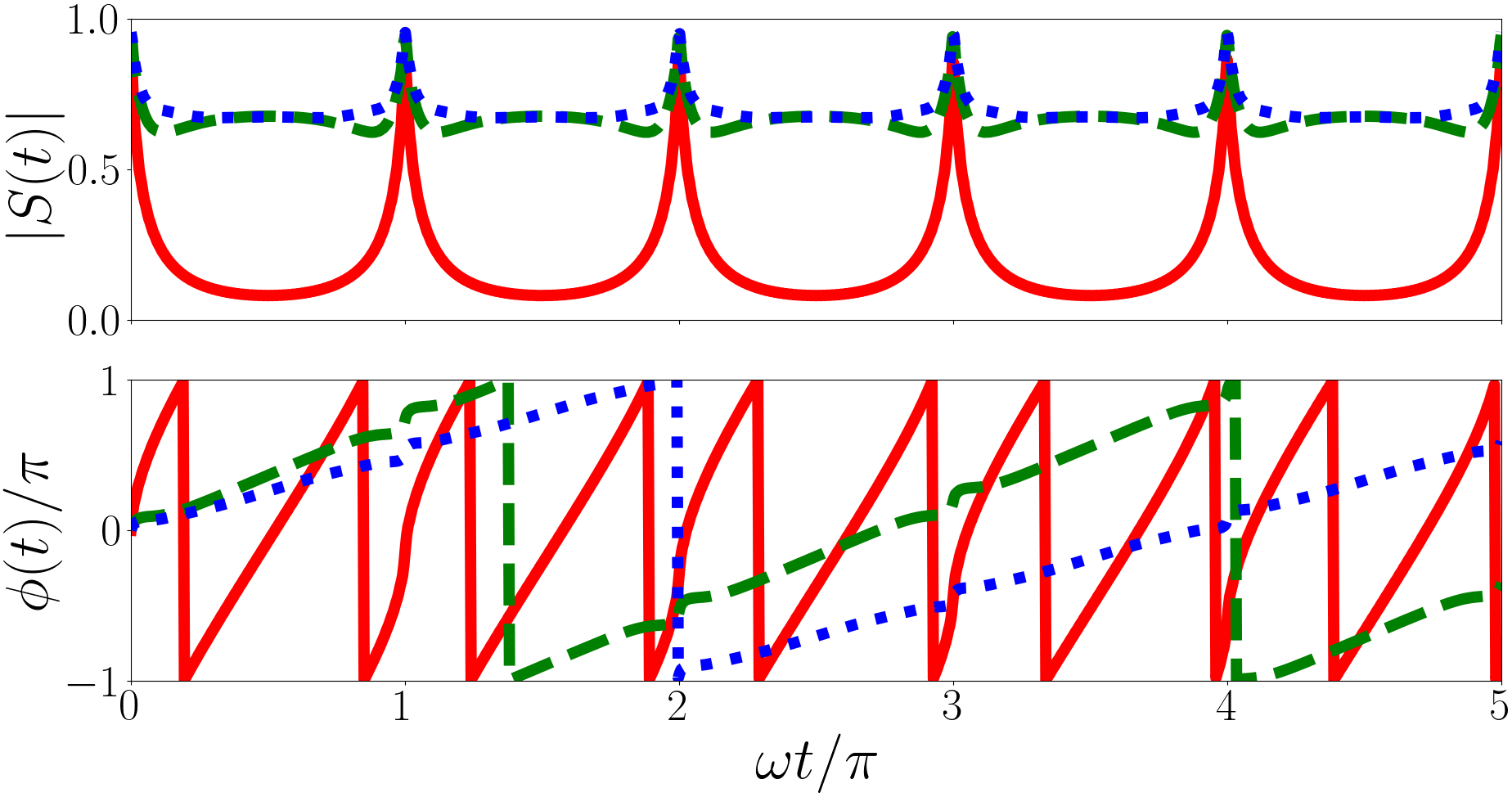}
\includegraphics[height=5.5cm, width=8.5cm]{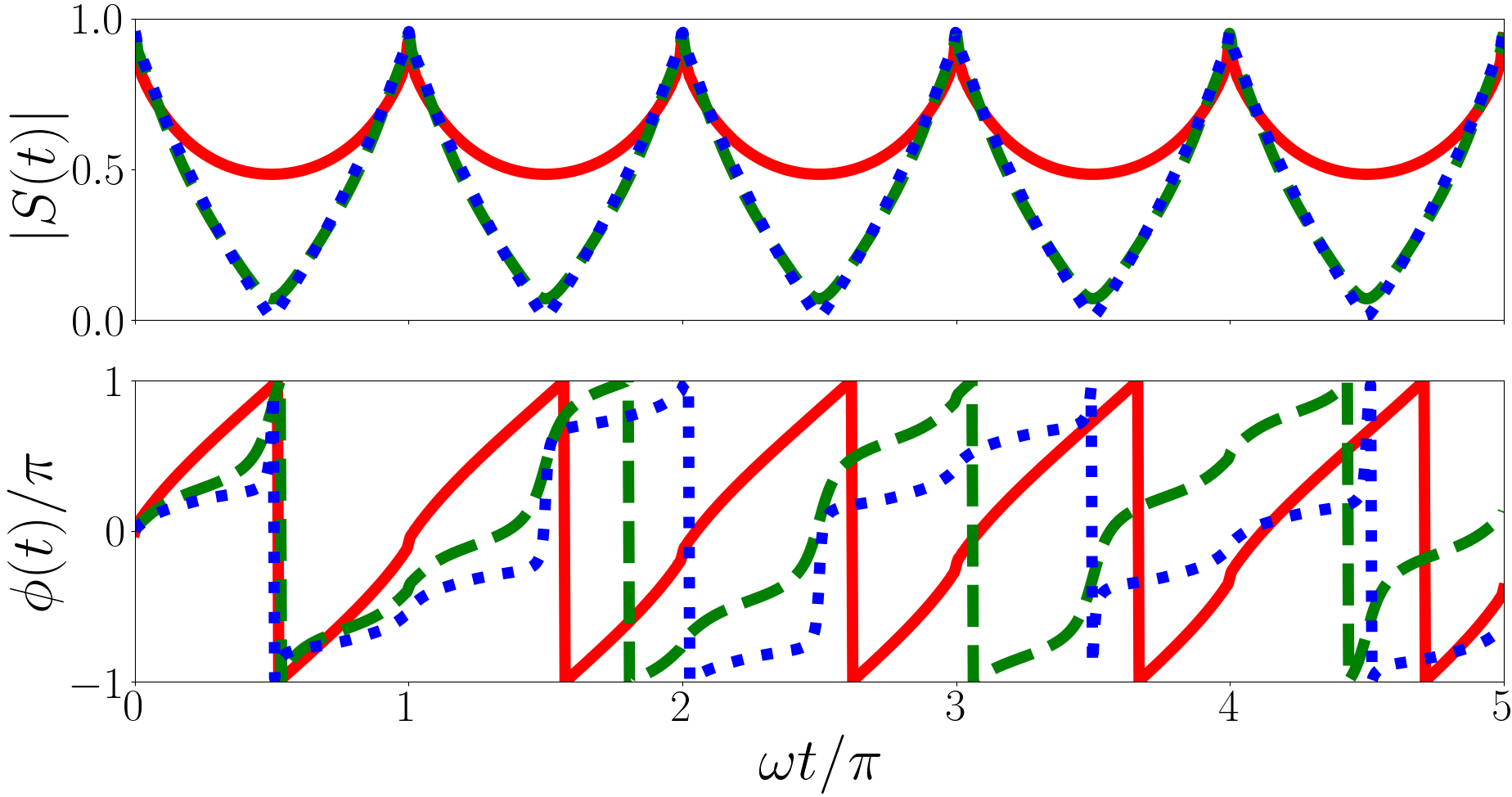}
\includegraphics[height=5.5cm, width=8.5cm]{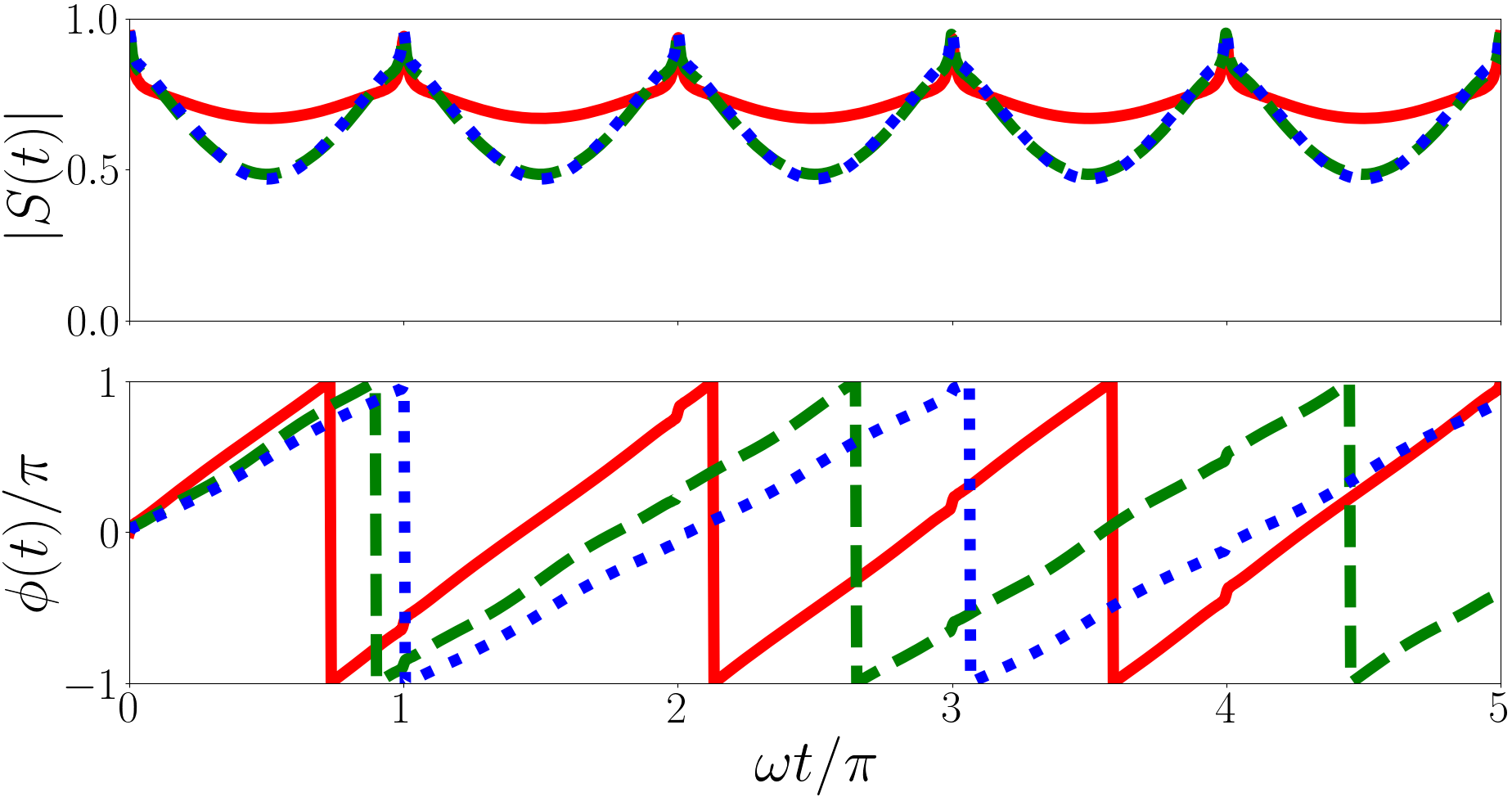}
\caption{Ramsey signal of the system quenched from unitarity to non-interacting. In each panel $s_{\rm i}=i\cdot1.006\dots$ and the solid red line corresponds to $q_{i}=0$, the dashed green line to $q_{i}=1$, and the dotted blue line to $q_{i}=2$. The upper panel uses $R_t=a_{\mu}$ to calculate the Efimov energy spectrum, and thus the energy of the initial state. The middle panel uses $R_t=5a_{\mu}$ and the lower $R_t=10a_{\mu}$. These Ramsey signals are evaluated using Eq. (\ref{eq:RamseyDefn}) with 40 terms in each of the sums, 1600 terms total. We find that the summation is convergent.}
\label{fig:RamseyBackwards}
\end{figure}

In Fig. \ref{fig:RamseyBackwards} we plot the Ramsey signal of the backwards quench for a system initially in an Efimov state for a variety of Efimov energies. Unlike in the forwards quench the magnitude of the Ramsey signal of the backwards quench is periodic. This is because the non-interacting eigenenergies are all odd integer multiples of $\hbar\omega$. The difference between the post-quench eigenenergies are even integers leading to the magnitude having period $\pi/\omega$. However the phase is dominated by the largest term in Eq. (\ref{eq:RamseySignal}) and the angular frequencies of each term are irrational because the initial state is an Efimov state which, in general, has an irrational energy. This leads to the irregularity in the phase.

For $R_{t}=a_{\mu}$, $(E_{q_{\rm i}=0},s_{\rm i})\approx(-0.850,i\cdot1.006)$ (solid red line of the upper panel of Fig. \ref{fig:RamseyBackwards}) the largest terms are the overlaps with $(q,s)=(0,2)$ with square overlap $\approx0.14$ and $(q,s)=(1,2)$ with square overlap $\approx0.12$. These terms have periods of $\approx0.52\pi/\omega$ and $\approx 0.34\pi/\omega$. For $R_{t}=5a_{\mu}$ $(E_{q_{\rm i}=0},s_{\rm i})\approx(1.077,i\cdot1.006)$ (solid red line of the middle panel of Fig. \ref{fig:RamseyBackwards}) the largest term is the overlaps with $(q,s)=(0,2)$ with square overlap $\approx0.58$ and period $\approx\pi/\omega$. For $R_{t}=10a_{\mu}$ $(E_{q_{\rm i}=0},s_{\rm i})\approx(1.602,i\cdot1.006)$ (solid red line of the lower panel of Fig. \ref{fig:RamseyBackwards}) the largest term is the overlaps with $(q,s)=(0,2)$ with square overlap $\approx 0.72$ and period $\approx1.43\pi/\omega$.
 
\subsection{Particle separation}
\label{sec:ParticleSep}

We are not limited to calculating only the Ramsey signal. It is also possible to calculate the particle separation, $\langle \tilde{R}(t) \rangle$.

The expectation value of $\tilde{R}(t)$ is given
\begin{eqnarray}
\langle \tilde{R}(t)\rangle = \bra{\Psi'(t)}\tilde{R}\ket{\Psi'(t)} &=& \sum_{j,j'}\bra{\Psi_{\rm i}(0)}\ket{\Psi'_{j}}\bra{\Psi'_{j'}}\ket{\Psi_{\rm i}(0)}\nonumber\\
&\:&\times \bra{\Psi'_{j}}\tilde{R}\ket{\Psi'_{j'}} e^{-i(E_{j'}-E_{j})t/\hbar},\nonumber\\
\end{eqnarray}
where $\Psi_{\rm i}$ is the initial pre-quench state with energy $E_{\rm i}$ and $\Psi'(t)$ is the post-quench state. $\Psi'_{j}$ and $\Psi'_{j'}$ are eigenstates of the post-quench system with eigenenergy $E_{j}$ and $E_{j'}$ respectively, with the sums	 over $j$ and $j'$ taken over all post-quench eigenstates.

The COM wavefunction is independent of the interparticle interaction and does not impact the post-quench dynamics. Due to the hyperangular wavefunction's orthogonality in $s$, two sums over $s$ and $s'$ collapse into a single sum over $s$. Hence $\langle \tilde{R}(t) \rangle$ is given
\begin{eqnarray}
\langle \tilde{R}(t) \rangle &=& \sum_{q',q}\sum_{s}\bra{F_{q_{\rm i}s_{\rm i}}\phi_{s_{\rm i}}}\ket{F_{q's}\phi_{s}}\bra{F_{qs}\phi_{s}}\ket{F_{q_{\rm i}s_{\rm i}}\phi_{s_{\rm i}}}\nonumber\\
&\;& \times \bra{F_{q's}\phi_{s}}\tilde{R}\ket{F_{qs}\phi_{s}}e^{-i(E_{qs}-E_{q's})t/\hbar},\label{eq:ExpectR}
\end{eqnarray}
As in Eq. (\ref{eq:RamseySignal}) indices with subscript i refer to the initial state and indices with no subscript refer to the post-quench eigenstates. As with the Ramsey signal all relevant integrals are presented in the appendix.

\begin{figure}
\includegraphics[height=5.4cm, width=8.5cm]{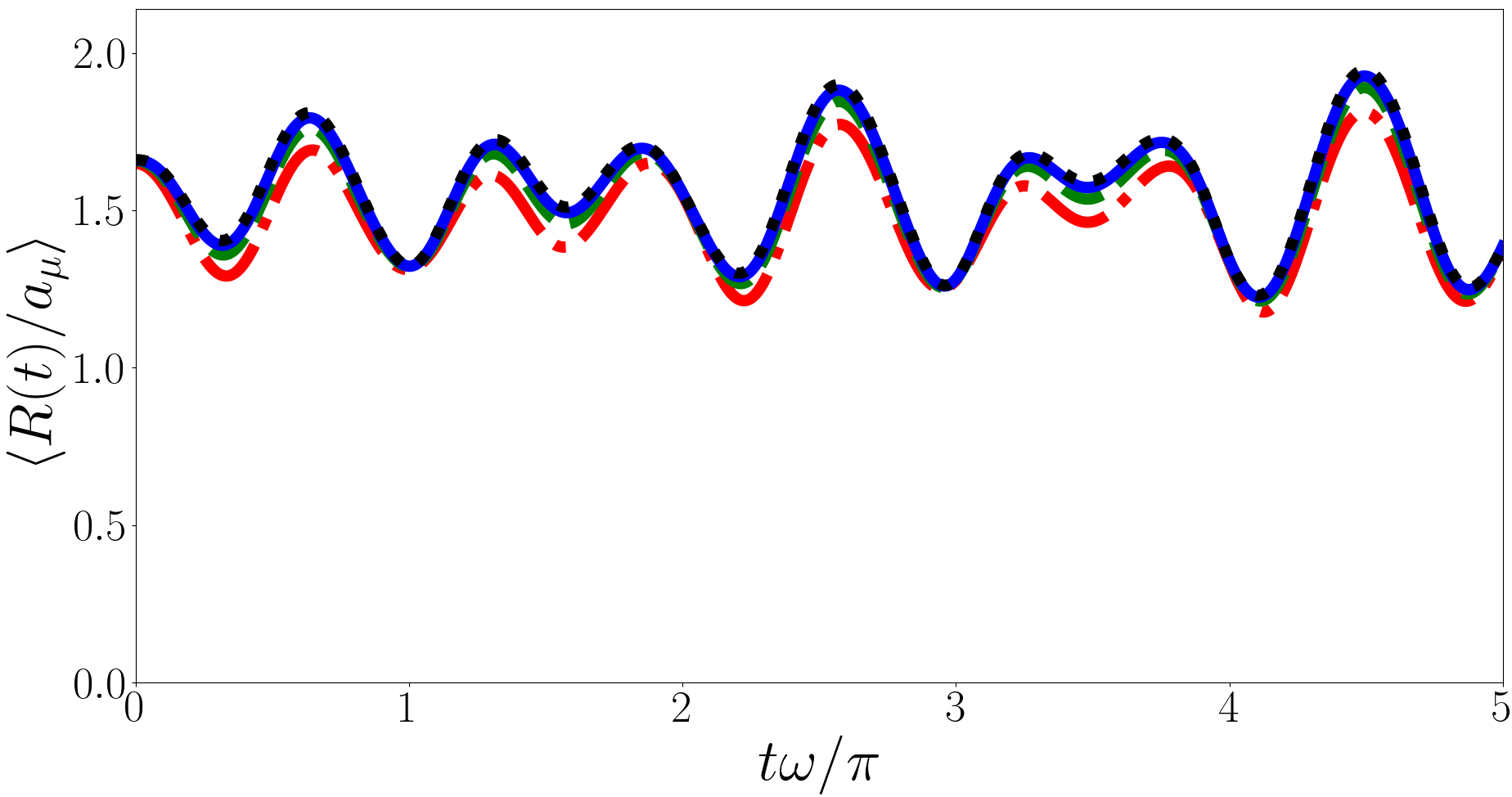}
\includegraphics[height=5.4cm, width=8.5cm]{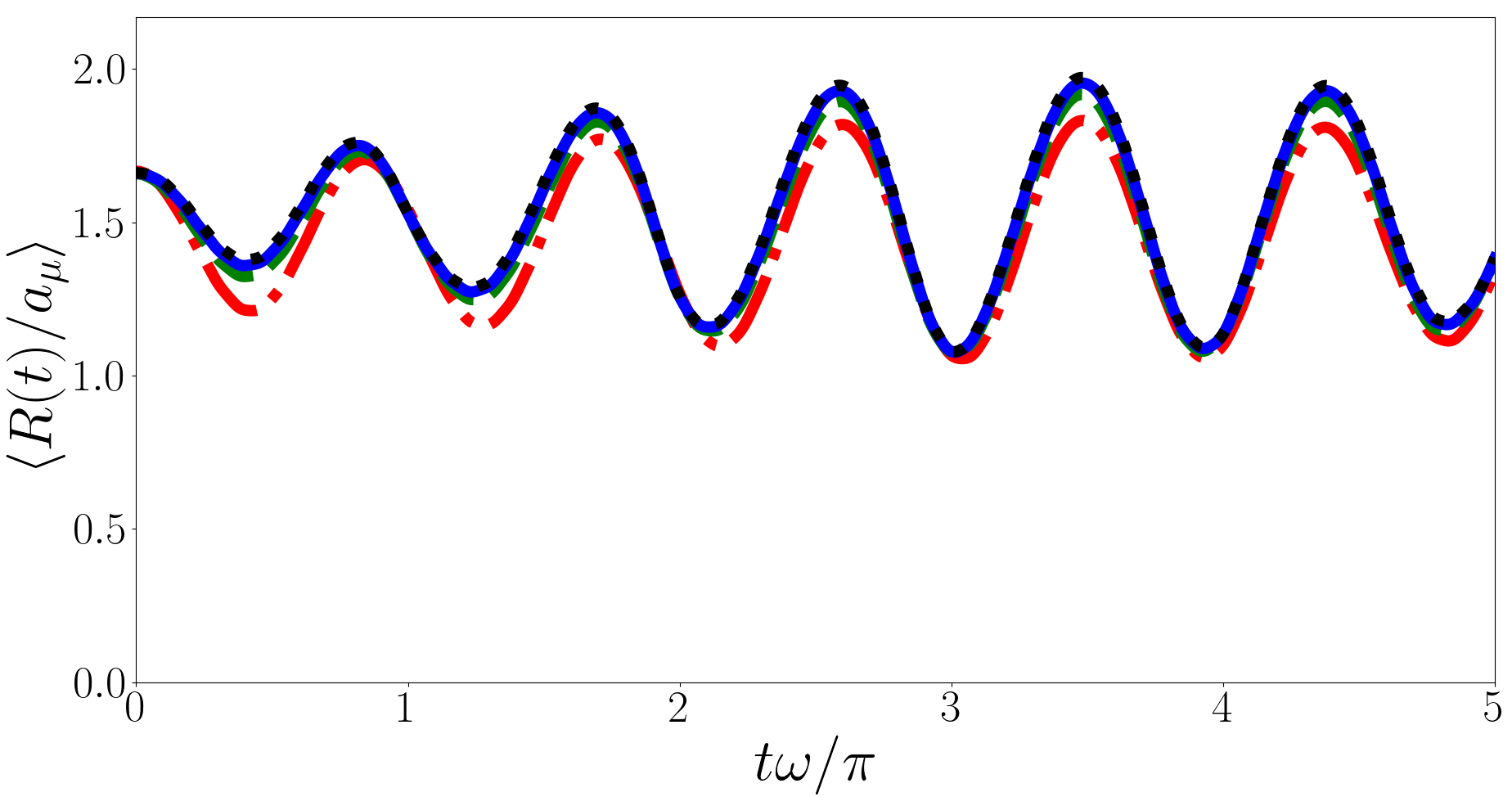}
\includegraphics[height=5.4cm, width=8.5cm]{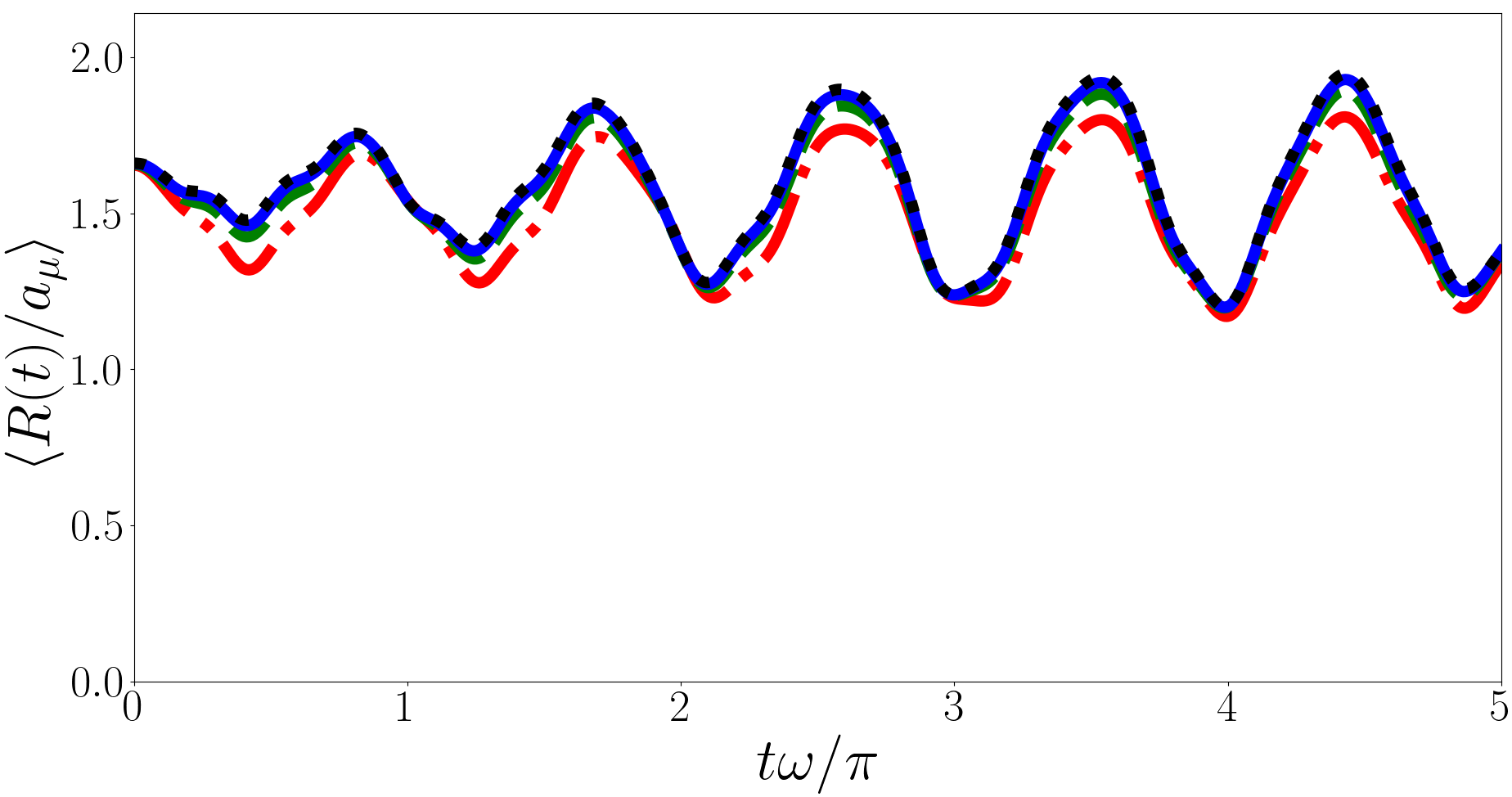}
\caption{$\langle \tilde{R}(t) \rangle$ of a system initially in the ground state quenched from non-interacting to unitarity. The upper panel corresponds to $R_{t}=a_{\mu}$, the middle to $R_{t}=5a_{\mu}$ and the lower to $R_{t}=10a_{\mu}$. Efimov states with $q\geq-1$ are included in the calculation except $R_{t}=a_{\mu}$ where $E_{q=-1}\approx-566\hbar\omega$ does not contribute meaningfully. The dot-dashed red line corresponds to $N_{\rm max}=3$, the dashed green line to $N_{\rm max}=6$, the solid blue line to $N_{\rm max}=12$ and the dotted black line to $N_{\rm max}=24$. We find that the summation is convergent.}
\label{fig:ExpectRForwards}
\end{figure}

In Fig. \ref{fig:ExpectRForwards} we plot $\langle \tilde{R}(t) \rangle$ for a system initially in the non-interacting ground state quenched to unitarity with the upper, middle and lower panels corresponding to $R_{t}=a_{\mu}$, $R_{t}=5a_{\mu}$, $R_{t}=10a_{\mu}$ respectively. For $R_{t}\geq5a_{\mu}$ we include states with $q\geq-1$ in the calculations but for $R_{t}=a_{\mu}$ $E_{q=-1}\approx-566\hbar\omega$ and this state does not meaningfully contribute so we only include the $q\geq0$ Efimov states in the calculation. In Eq. (\ref{eq:ExpectR}) terms with $q=q'$ are constants and the $s$ contributions to the energies cancel out, the angular frequencies depend only on $q$ and $q'$. The universal state terms oscillate with an angular frequency that is an even integer multiple of $\omega$ because $q-q'$ is an integer but the Efimov state terms oscillate with irrational angular frequencies because the differences between the Efimov energies, $E_{q}$ and $E_{q'}$, are irrational in general. For each plot we have calculated $\langle \tilde{R}(t) \rangle$ by summing up to $N_{\rm max}=3,6,12,24$ terms in each of the three sums in Eq. (\ref{eq:ExpectR}), and we find that the sum is convergent.

For $R_{t}=a_{\mu}$ the largest oscillating terms are $(q',q,s)=(q,q',s)=(0,1,i\cdot1.006\dots)$ with total coefficient $\approx0.17$ and $E_{q=1}-E_{q=0}\approx3.12\hbar\omega$ and $(q,q',s)=(1,2,i\cdot1.006\dots)$ with total coefficient $\approx0.09$ and $E_{q=2}-E_{q=1}\approx2.12\hbar\omega$ . This implies characteristic periods of $\approx2\pi/3\omega$ and $\approx\pi/\omega$. For $R_{t}=5a_{\mu}$ the largest oscillating terms are $(q,q',s)=(0,1,i\cdot1.006\dots)$ with total coefficient $\approx0.3$ and $E_{q=1}-E_{q=0}\approx 2.3\hbar\omega$ and $(q,q',s)=(0,1,4.465)$ with total coefficient $\approx0.054$ and associated energy difference $2\hbar\omega$. This leads to characteristic periods of $\approx0.9\pi/\omega$ and $\pi/\omega$. For $R_{t}=10a_{\mu}$ the largest oscillating terms are $(q,q',s)=(0,1,i\cdot 1.006\dots)$ with total coefficient $\approx 0.22$ and $E_{q=1}-E_{q=0}\approx 2.3\hbar\omega$ and $(q,q',s)=(0,1,4.465)$ with total coefficient $\approx0.054$ and associated energy difference $2\hbar\omega$. This leads to characteristic periods of $\approx0.9\pi/\omega$ and $\pi/\omega$.

\begin{figure}
\includegraphics[height=5.4cm, width=8.5cm]{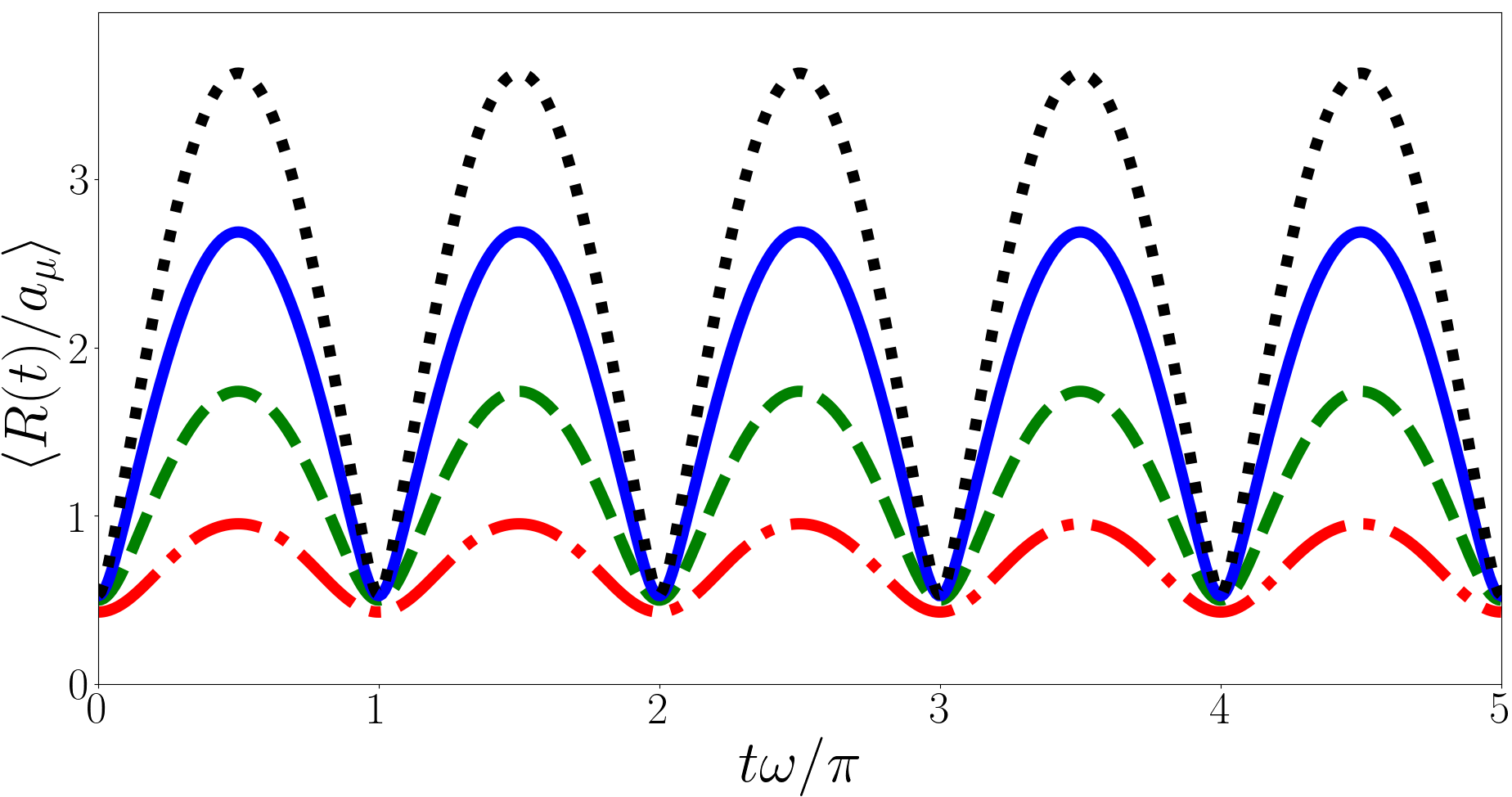}
\includegraphics[height=5.4cm, width=8.5cm]{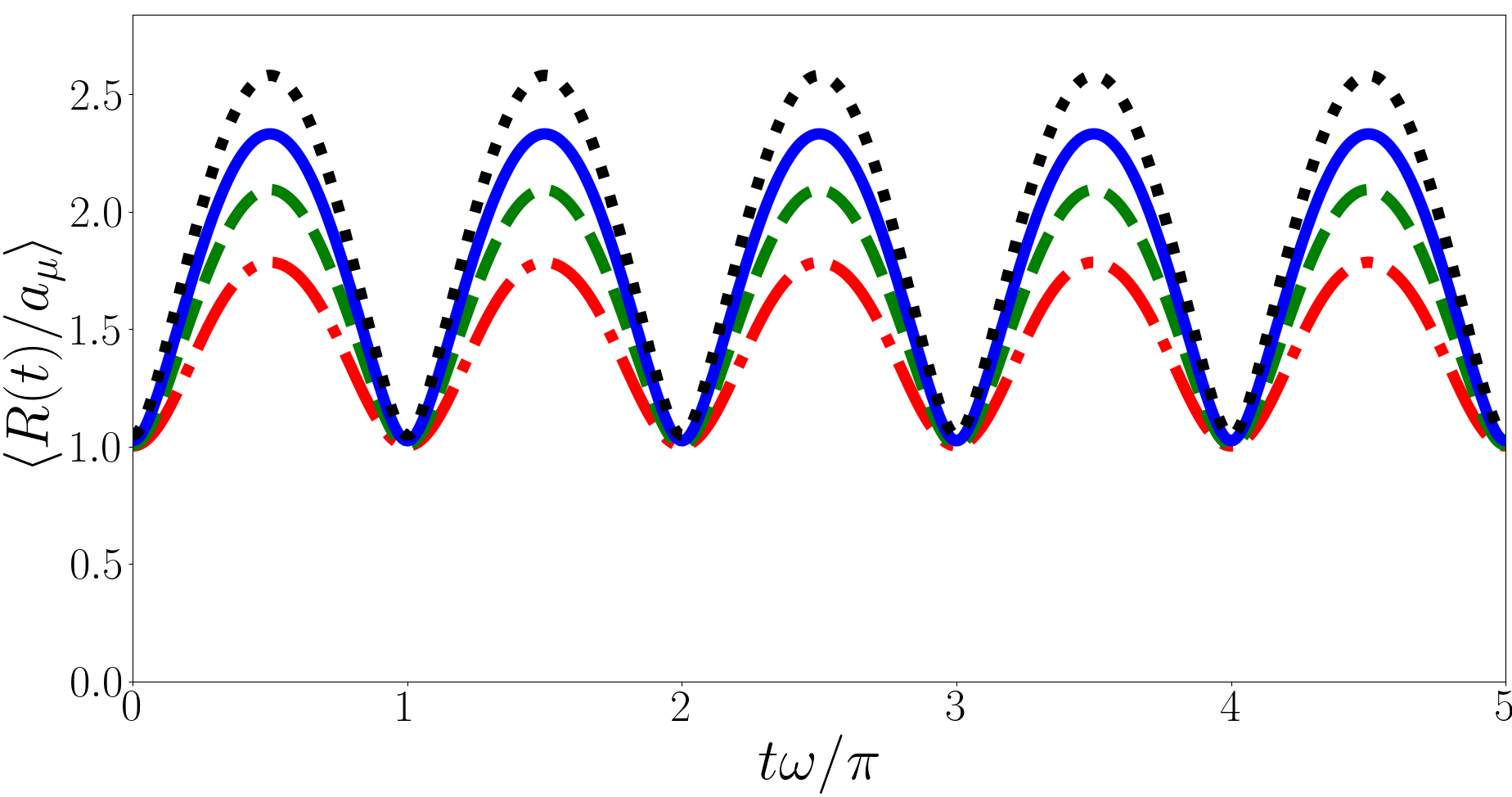}
\includegraphics[height=5.4cm, width=8.5cm]{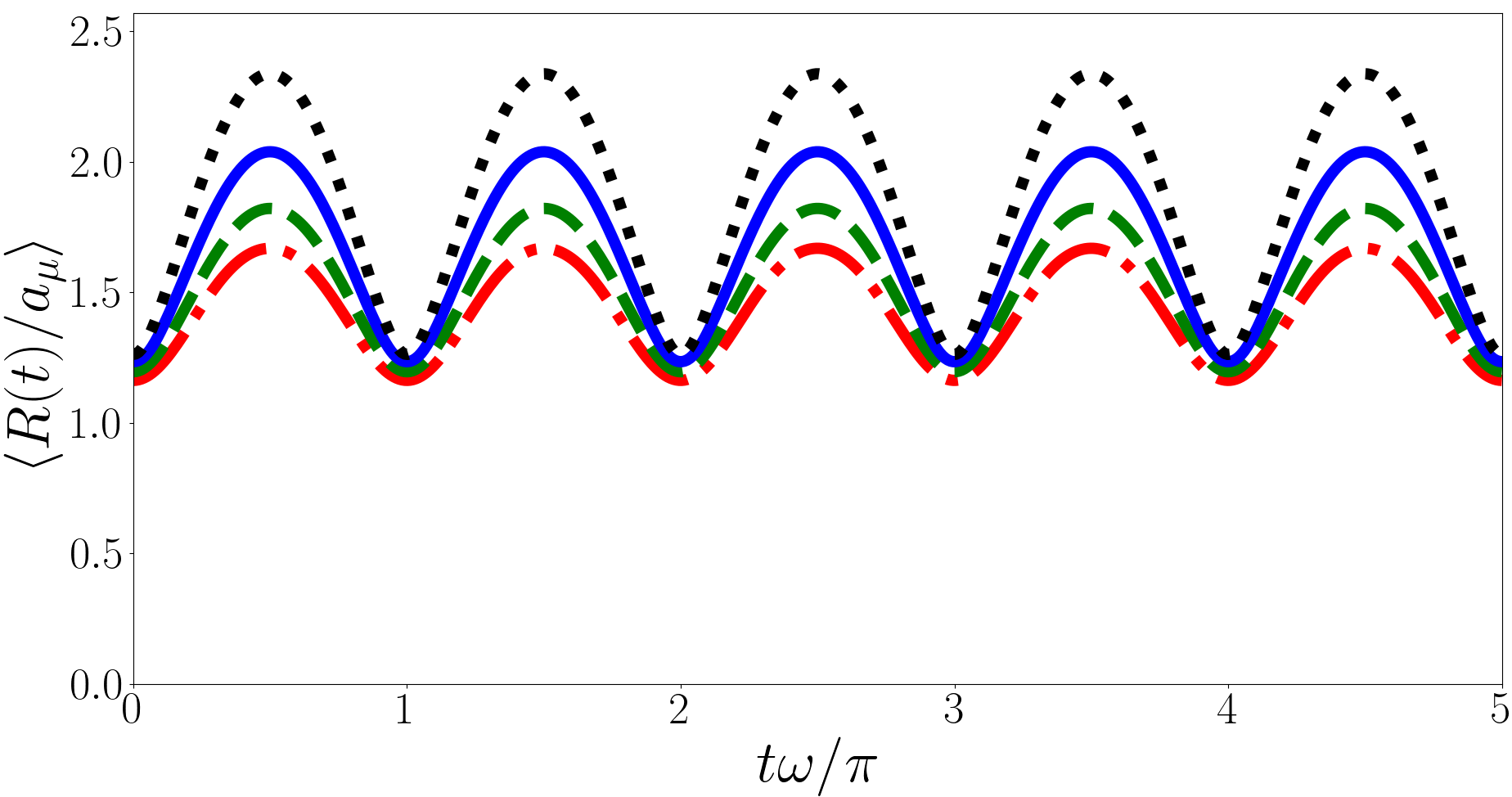}
\caption{$\langle \tilde{R}(t) \rangle$ of a system following a quench from unitarity to non-interacting. The upper, middle and lower panels correspond to $R_{t}=a_{\mu}$, $R_{t}=5a_{\mu}$ and $R_{t}=10a_{\mu}$ and the corresponding initial states are Efimov states with $q=0$. The dot-dashed red line corresponds to $N_{\rm max}=3$, the dashed green line to $N_{\rm max}=6$, the solid blue line to $N_{\rm max}=12$ and the dotted black line to $N_{\rm max}=24$.}
\label{fig:ExpectRBackwards}
\end{figure}

In Fig. \ref{fig:ExpectRBackwards} we plot $\langle \tilde{R}(t) \rangle$ for the backwards quench where the system is initially in a variety of Efimov states. Unlike the forwards quench we find that $\langle \tilde{R}(t) \rangle$ is periodic for the backwards quench. This is because in the backwards quench the post-quench states are universal states where the differences between eigenenergies are always even multiples of $\hbar\omega$, leading to a period of $\pi/\omega$. However similar to how Ref. \cite{kerin2020two} found a divergence in $r=|\vec{r}_{2}-\vec{r}_{1}|$ in the backwards quench we find that $\langle \tilde{R}(t) \rangle$ also diverges for the backwards quench. In particular we find that it is logarithmically divergent with the number of terms in the summation, i.e. $\langle \tilde{R}(t\neq n\pi/\omega) \rangle\propto\ln(N_{\rm max})$. This divergence is not exclusively due to the Efimov states as the divergence is present even when there are no Efimov states \cite{kerin2022quench}.

This divergence is quite unusual, it is not obvious why it occurs nor why it occurs only for the reverse quench. To investigate further we look at how the probability distribution of $R$, $P(R,t)$, evolves over time for both quenches. $P(R,t)$ is given
\begin{eqnarray}
P(R',t)=\bra{\Psi'(t)}\delta(R'-R)\ket{\Psi'(t)},
\label{eq:ProbDistrib}
\end{eqnarray}
where $\Psi'(t)$ is the post-quench state.

\begin{figure}
\includegraphics[height=5.4cm, width=6cm]{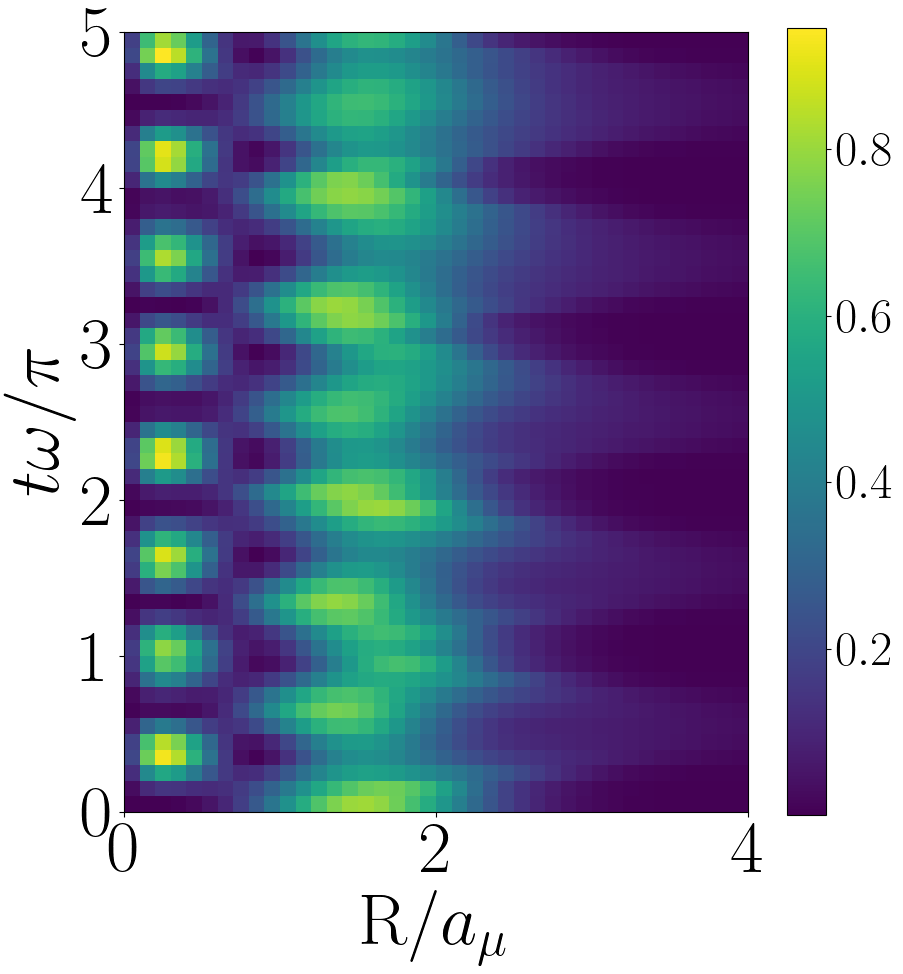}
\includegraphics[height=5.4cm, width=6cm]{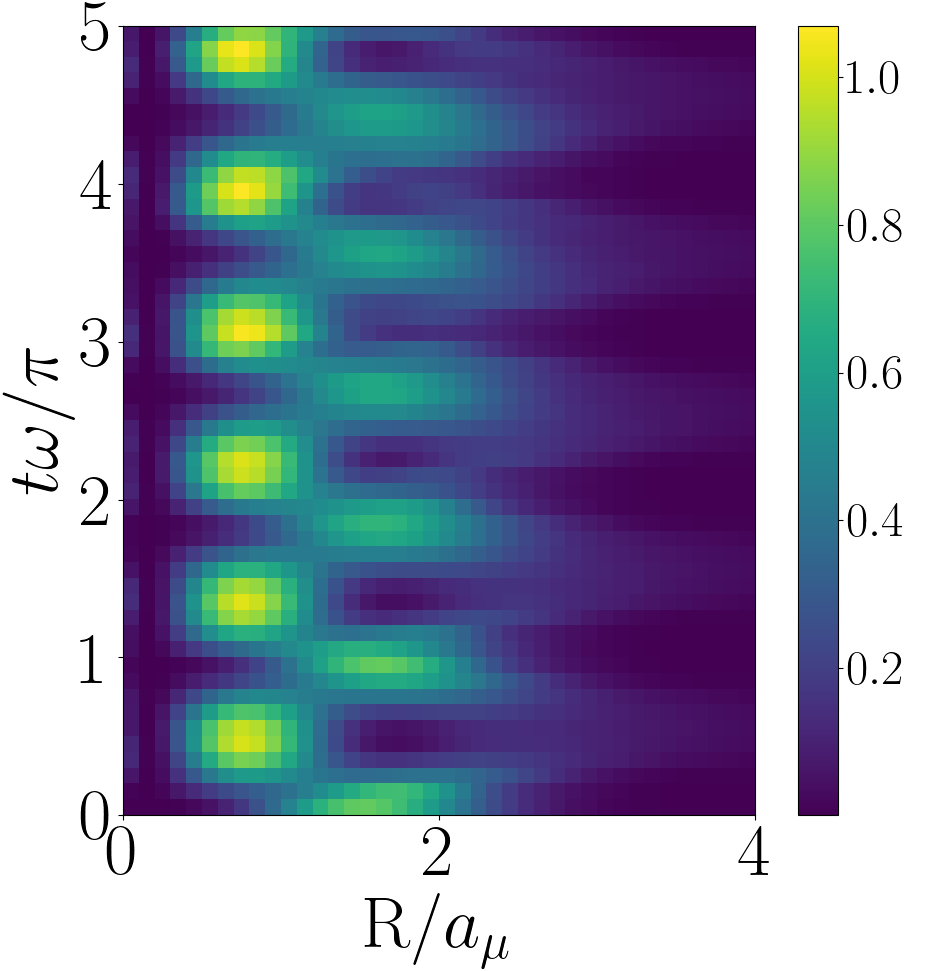}
\includegraphics[height=5.4cm, width=6cm]{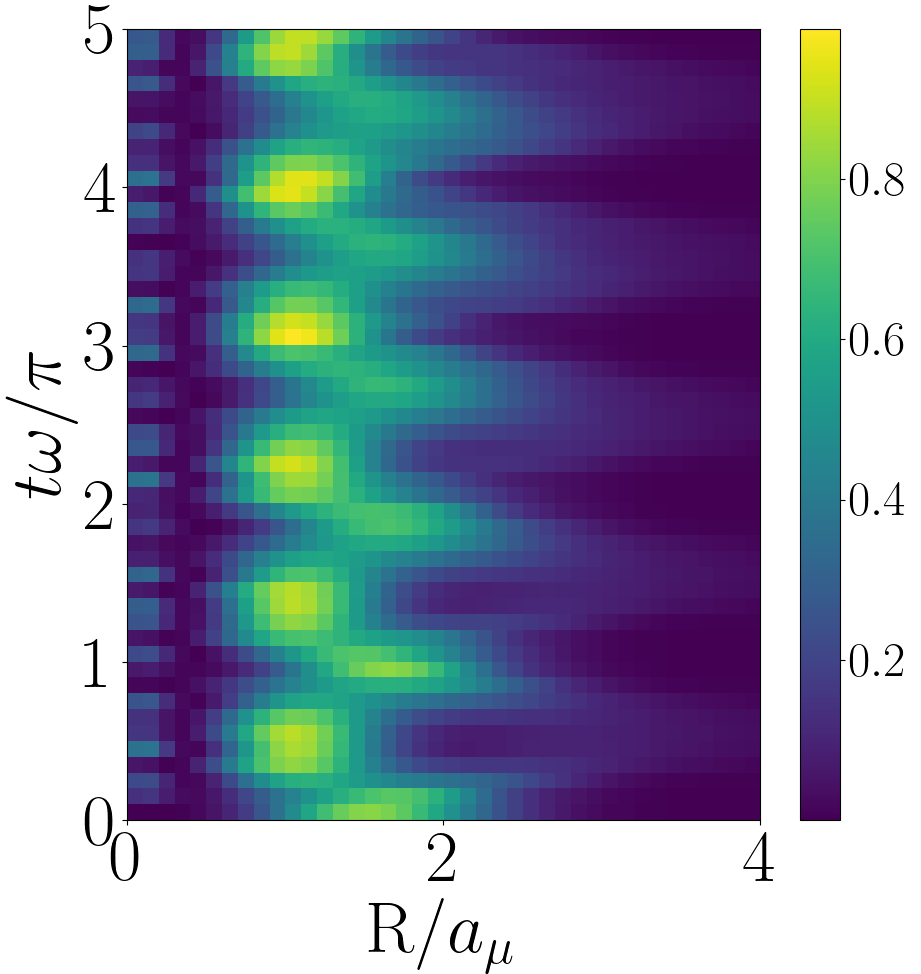}
\caption{
The evolution of the probability distribution of the hyperradius, Eq. (\ref{eq:ProbDistrib}), for the forwards quench. The upper, middle and lower panels correspond to $R_{t}/a_{\mu}=1,5$ and $10$ respectively. The horizontal axis is the hyperradius and the vertical axis is time with dark blue corresponding to low probability density, and yellow to high density. For all plots the initial state is $(q_{\rm i},s_{\rm i})=(0,2)$ and each plot is constructed with $N_{\rm max}=24$. In this case $P(R,t)$ is convergent with $N_{\rm max}$.
}
\label{fig:PRtForwardEvolution}
\end{figure}

In Fig. \ref{fig:PRtForwardEvolution} we plot the evolution of $P(R,t)$ for the forwards quench with $R_{t}/a_{\mu}=1,5$ and $10$ in the upper, middle and lower panels respectively. For all values of $R_{t}$ we see a qualitatively similar evolution, the system oscillates between a broad distribution and a tightly peaked one with a smaller mean value. This oscillation is only approximately periodic due to the influence of the irrational Efimov energies. The broad distribution corresponds to the initial universal state and the tightly peaked distribution is dominated by the Efimov states, the system oscillates between these two regimes.

To understand this oscillation it is useful to consider $\langle \tilde{R} \rangle$ for the initial and post-quench states. For example for $R_{t}=a_{\mu}$ the states with the largest overlaps with the initial are $(E_{q=1},s)\approx(2.27\hbar\omega,i \cdot 1.006)$ with a square overlap of $\approx 0.666$ and $(E_{q=0},s)\approx (-0.85\hbar\omega,i \cdot 1.006)$ with square overlap $\approx0.14$, these states have $\langle \tilde{R} \rangle\approx 1.5$ and $\approx 0.57$ respectively. The initial state is $(q,s)=(0,2)$ so we have $\langle \tilde{R}(t=0) \rangle\approx 1.66$, hence the position of the initial broad distribution is to the right of the tightly peaked Efimov distribution. As $R_{t}$ increases $\langle \tilde{R} \rangle$ of the strongly overlapping Efimov states increases but they are, on average, still less than $\langle \tilde{R} \rangle$ of the initial universal state, hence the peak of the narrow distribution moves rightward with increasing $R_{t}$. Additionally as $R_{t}$ increases the narrow Efimov distribution broadens because the higher energy Efimov states are simply broader. Note in the third panel of Fig. \ref{fig:PRtForwardEvolution} one can see small local peaks in probability near $\tilde{R}=0$. These come from the $(E_{q=-1},s)\approx(-5.6,i \cdot 1.006)$ state and have a square overlap with the initial state of $\approx 0.01$ and $\langle \tilde{R} \rangle\approx 0.3$, the $(E_{q=-1},s)$ state is also accounted for in the $R_{t}=5a_{\mu}$ calculation but the overlap is approximately 50 times smaller.

\begin{figure}
\includegraphics[height=5.4cm, width=8.5cm]{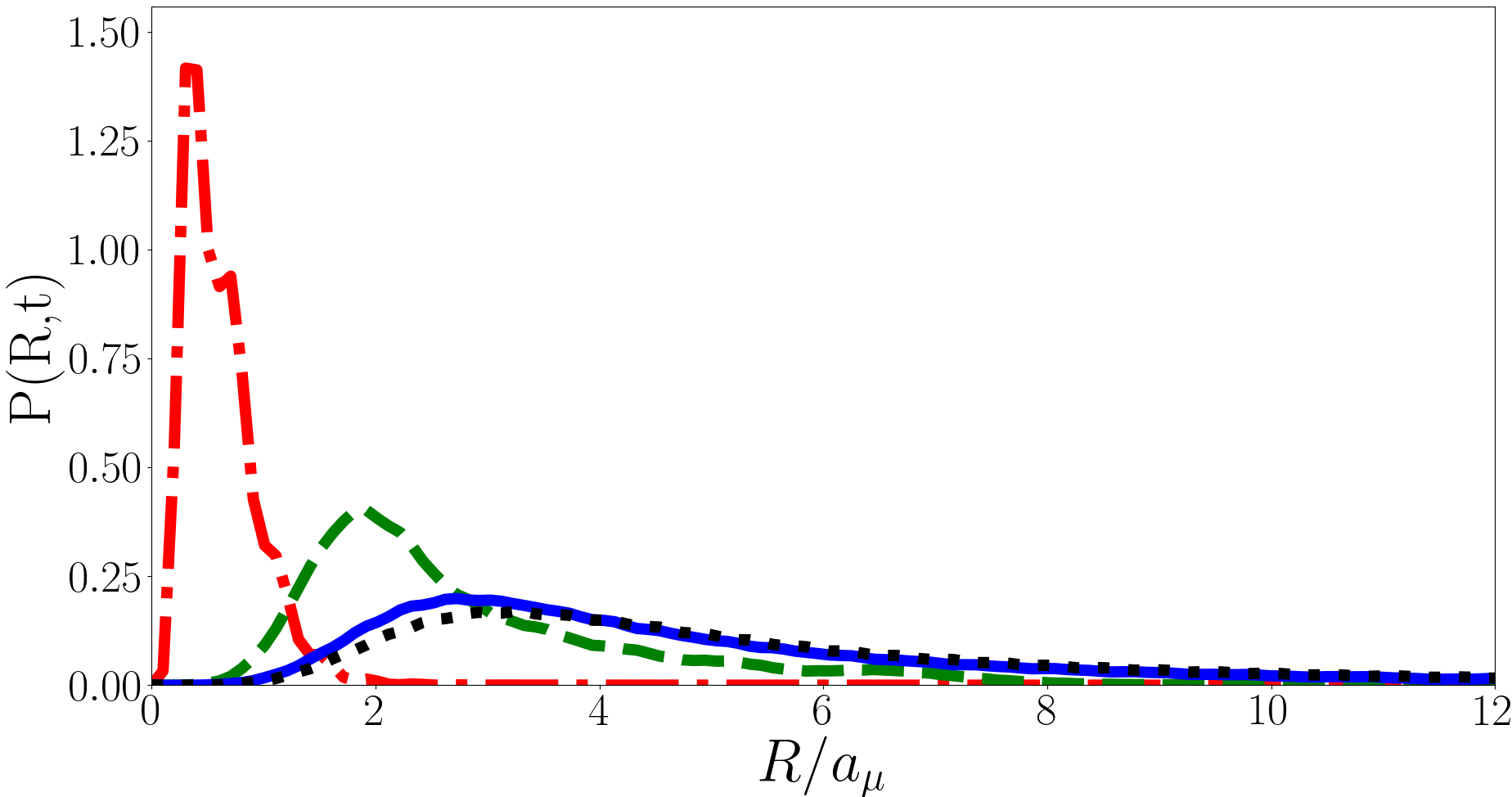}
\includegraphics[height=5.4cm, width=8.5cm]{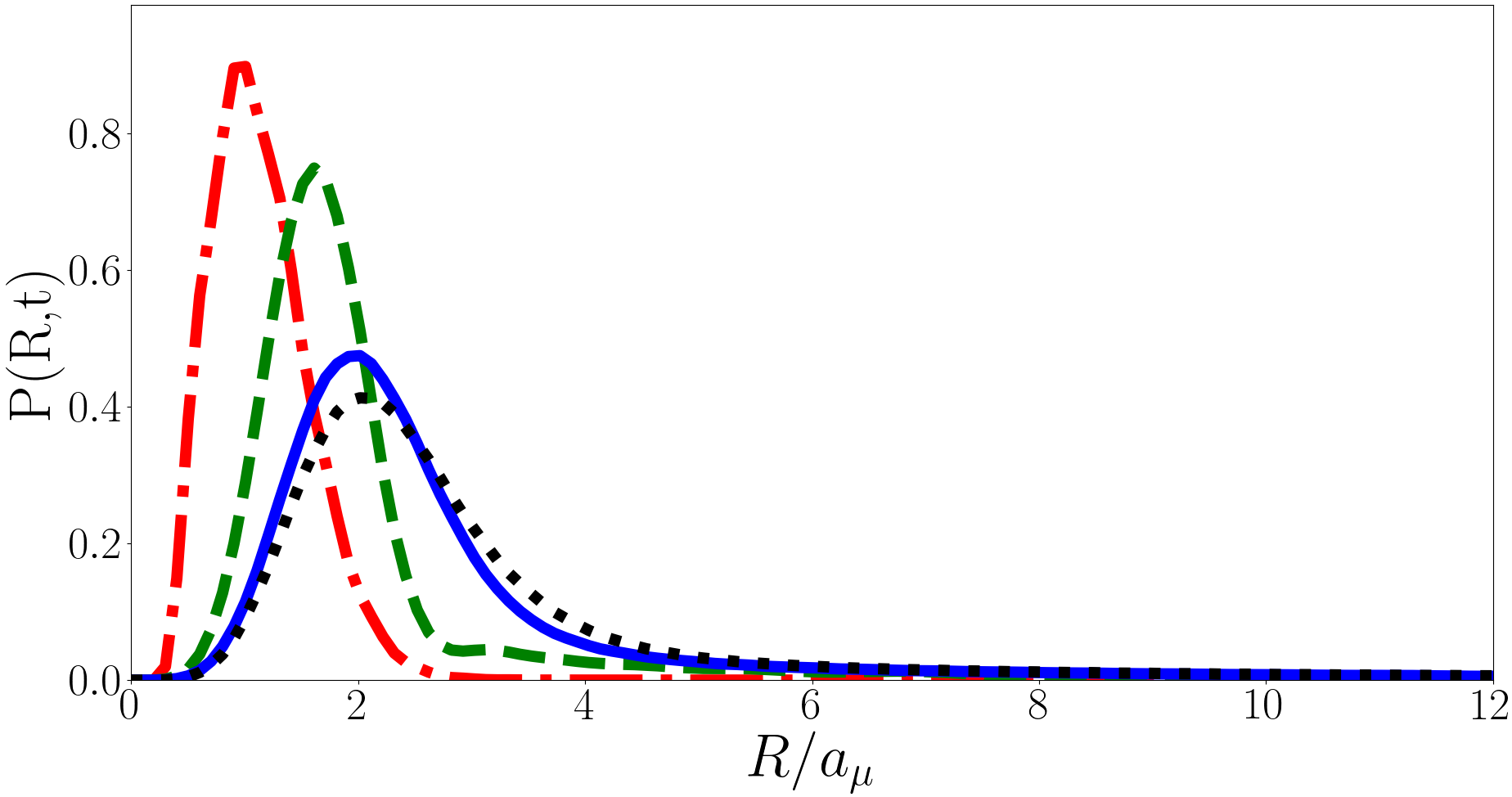}
\includegraphics[height=5.4cm, width=8.5cm]{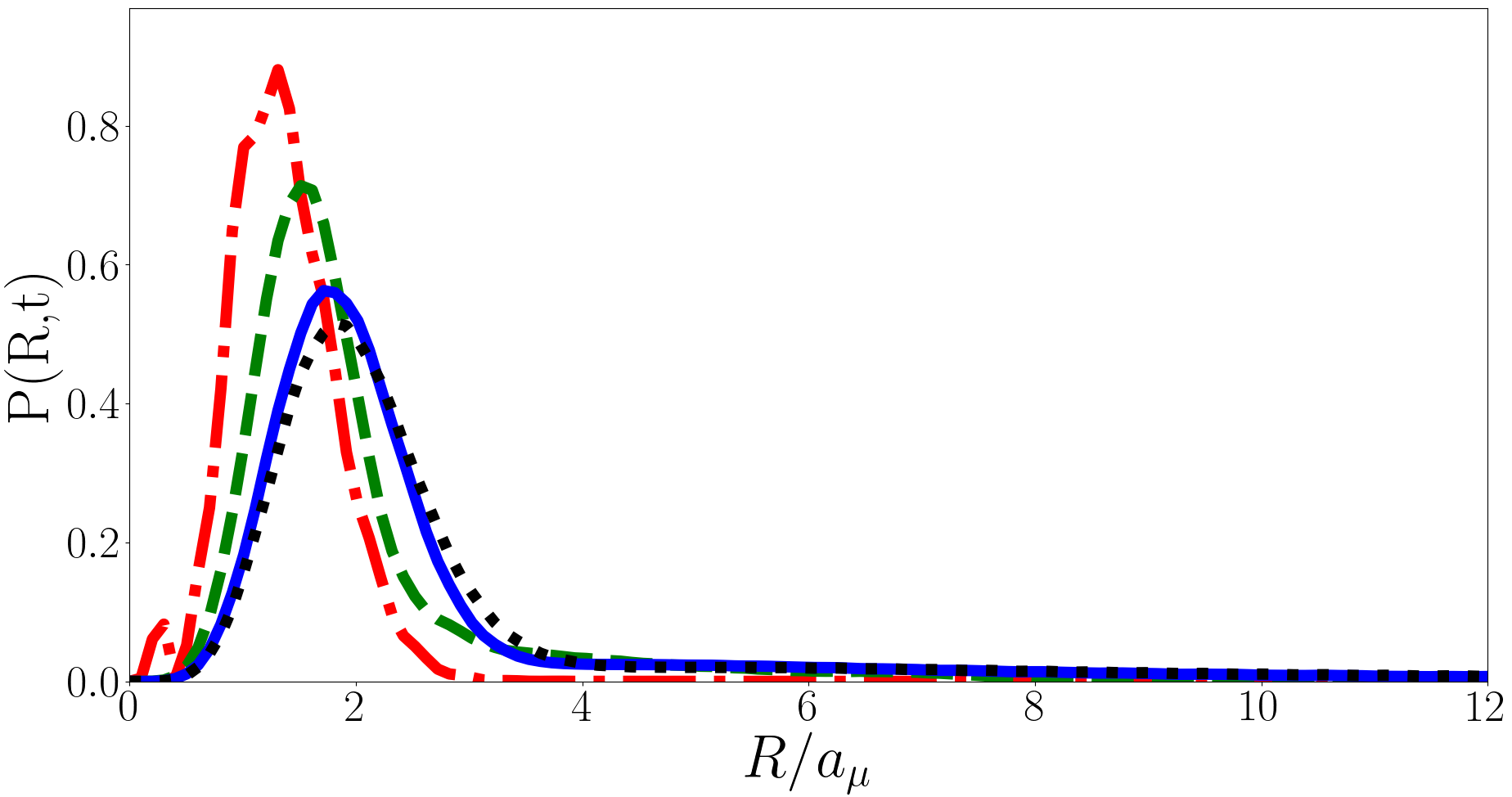}
\caption{
The evolution of the probability distribution of the hyperradius, Eq. (\ref{eq:ProbDistrib}), for the backward quench plotted at $t=0$ (dot-dashed red line), $t=0.17\pi/\omega$ (dashed green), $t=0.34\pi/\omega$ (solid blue) and $t=\pi/\omega$ (dotted black). The initial states in the upper, middle and lower panels are the $q=0$ Efimov states for $R_{t}/a_{\mu}=1,5$ and $10$ respectively. All calculations are performed with $N_{\rm max}=60$, unlike in the forwards quench we find that $P(R,t)$ is only convergent for $t=0$.
}
\label{fig:PRtBackwardEvolution}
\end{figure}

\begin{figure}
\includegraphics[height=5.4cm, width=8.5cm]{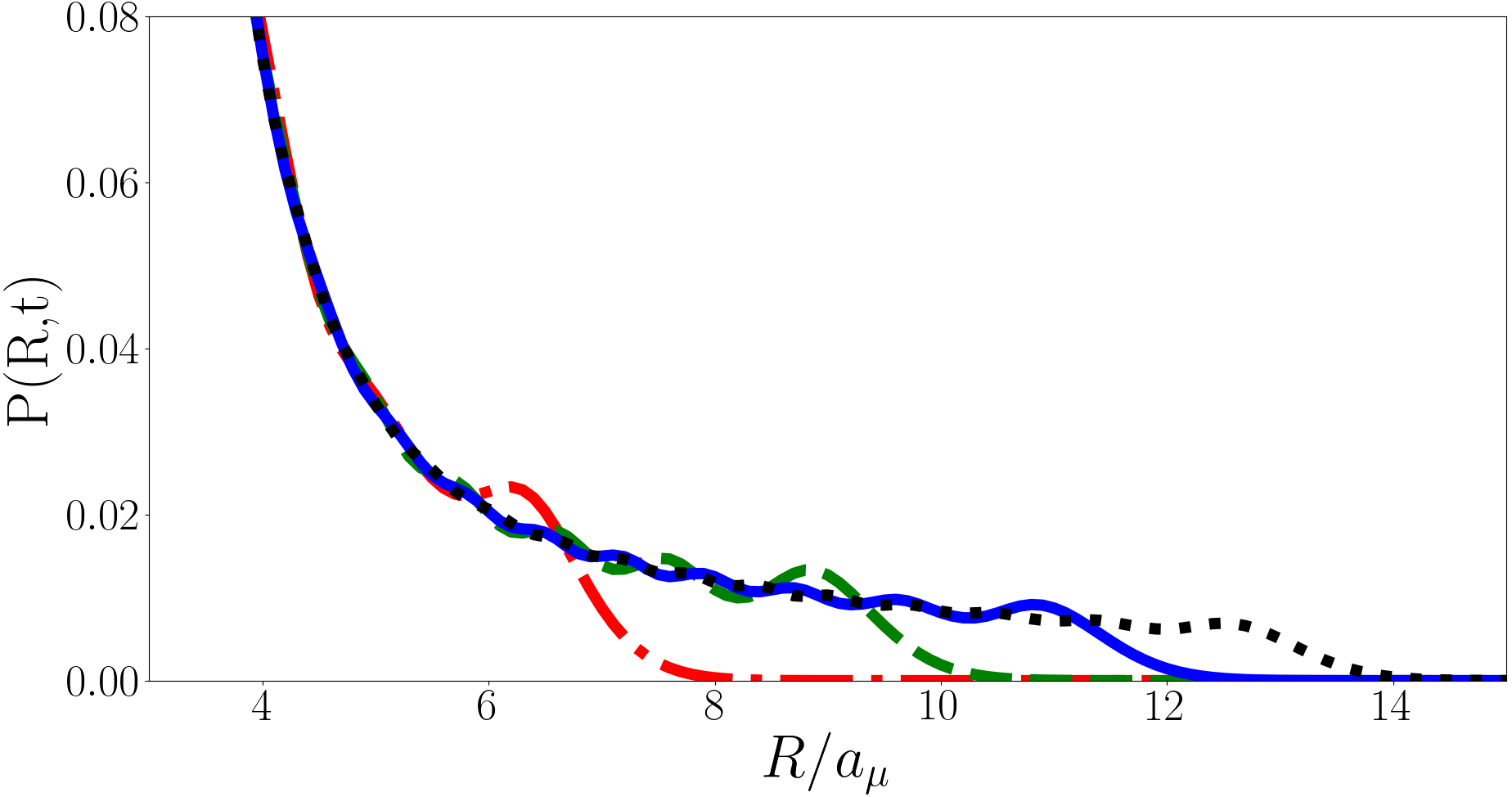}
\caption{
The tail of $P(R,t=\pi/2\omega)$ for the reverse quench with the energy of the initial Efimov state given $E\approx1.077$ ($q=0$ for $R_{t}=5a_{\mu}$) for various values of $N_{\rm max}$. The dot-dashed red line corresponds to $N_{\rm max}=10$, the dashed green line to $N_{\rm max}=20$, the solid blue line to $N_{\rm max}=30$ and the dotted black line to $N_{\rm max}=40$.
}
\label{fig:PRtBackwardTail}
\end{figure}

In Fig, \ref{fig:PRtBackwardEvolution} we plot $P(R,t)$ for the backwards quench at $t=0,0.17\pi/\omega,0.34\pi/\omega$ and $\pi/2\omega$. Unlike the forwards quench the evolution of $P(R,t)$ here is periodic. The mean of $P(R,t)$ increases with time, reaching a maximum at $\pi/2\omega$ before returning to its initial shape at $t=\pi/\omega$, this then repeats with period $\pi/\omega$. Initially $P(R,t)$ is tightly peaked but develops a long tail as it evolves and in Fig. \ref{fig:PRtBackwardTail} we present the long tail in detail for various values of $N_{\rm max}$. The tail behaves approximately like $\tilde{R}^{-2}$ until ending in an exponential-like ``cut-off''. This cut-off occurs at larger $R$ for larger $N_{\rm max}$ and in the limit of $N_{\rm max}\rightarrow \infty$ the behaviour of the tail of $P(R,t)$ approaches $\tilde{R}^{-2}$ with no cut-off. This means that the integral of $P(R,t)$ over $R$ from $R=0$ to $R\rightarrow\infty$ is finite and properly normalised in the $N_{\rm max}\rightarrow\infty$ limit, however $RP(R,t)$ has a $\tilde{R}^{-1}$ tail and so is the integral is not finite, hence the divergence in $\langle \tilde{R}(t) \rangle$ for the backwards quench.  

Physically speaking there are two likely candidates for the source of the divergence; the zero-range contact interaction and the instantaneous nature of the quench, in reality atoms interact at some finite range and the quench in $a_{\rm s}$ occurs over some finite time. These are two non-physical inputs into this model and may be responsible for the non-physical outputs.

By considering the finite range of the interaction it is possible to estimate a maximum value of $\langle \tilde{R}(t) \rangle$. The lengthscale of the interaction provides a justification for a maximum energy and thus a cut-off in Eq. (\ref{eq:ExpectR}). Specifically, the range of interaction defines a minimum de Broglie wavelength which defines a maximum energy and thus the cut-off. For sodium in a 1kHz trap and assuming a van der Waals range of one nanometre we obtain an energy of $E_{\rm rel}\approx 8.7\times 10^6 \hbar\omega$ and so we predict $\langle \tilde{R}(t) \rangle_{\rm max}\approx21$ for an initial Efimov energy of $E_{q=0}\approx-0.85$. This is an order of magnitude larger than the amplitude of oscillations when the system is quenched from the non-interacting to the strongly interacting regime. 

In contrast it is difficult to quantify the effects of a finite duration quench. In the formalism used here only quenches between the non-interacting and unitary regimes can be described, meaning a quench to or from the intermediate regime can't be elucidated. However in the two-body case a quench between any two scattering lengths can be considered \cite{kerin2020two} so the effects of a finite duration quench can be investigated.

\section{Conclusion}
In this paper we have examined the effects of different Efimov energy spectra on the time dependent post-quench dynamics of an interacting few-body system. This was done in the context of three interacting bosons in a spherically symmetric trap, where the contact interactions were quenched from the non-interacting regime to the strongly-interacting regime (forwards quench) and vice-versa (backwards quench). In each case we were able to evaluate the post quench dynamics of both the Ramsey signal and the expectation value of the hyperradius.

For the forwards quench we find an irregularly repeating signal for both the Ramsey signal and $\langle \tilde{R}(t) \rangle$. For the Ramsey signal this is due to both the Efimov energies and unitary $s$-eigenspectrum being irrational in general. In the case of the particle separation the contributions from $s$ cancel out and the irregularity is due to the irrationality of the Efimov energies. In both cases the results are convergent and well defined.

For the backwards quench the magnitude of the Ramsey signal and $\langle \tilde{R}(t) \rangle$ oscillate with period $\pi/\omega$. This is because the non-interacting $s$-eigenvalues are even integers and Efimov states are not present when $a_{\rm s}=0$. The phase of the Ramsey signal is still irregular due to the influence of the initial irrational Efimov energy. However we find, analogous to previous results \cite{kerin2020two}, that the particle separation diverges logarithmically. By enforcing a cut-off on Eq. (\ref{eq:ExpectR}) motivated by a minimum de Broglie wavelngth derived from the van der Waals range we expect a maximum $\langle \tilde{R}(t) \rangle\approx 21$. This estimate of the size of the oscillations is extremely large compared to the forwards quench case.

\section{Acknowledgements}
A.D.K. is supported by an Australian Government Research Training Program Scholarship and by the University of Melbourne.
 
With thanks to Victor Colussi for illuminating discussions regarding the evaluation of the hyperangular integral.

\onecolumngrid
\section*{Appendix}
\label{sec:Appendix}
In this work we calculate quench observables of systems where the wavefunctions of the pre- and post-quench systems are known. To obtain these observables we need to perform numerous integrals involving these wavefunctions. In this appendix we present those integrals. Firstly, the Jacobian in hyperspherical coordinates is given by
\begin{eqnarray}
dV=d\vec{r}_{1}d\vec{r}_{2}d\vec{r}_{3}=\frac{3\sqrt{3}}{32}R^5\sin^2(2\alpha)dR d\alpha d\vec{\Omega}_{r}d\vec{\Omega}_{\rho}d\vec{C},
\end{eqnarray}
and for convenience we define
\begin{eqnarray}
\bra{F_{qs}(R)}\ket{F_{q's'}(R)} &=& \int_{0}^{\infty} RF_{qs}(R)^{*}F_{q's'}(R)dR,\\
\bra{\phi_{s}(\alpha)}\ket{\phi_{s'}(\alpha)} &=& \int\int\int_{0}^{\pi/2} \phi_{s}(\alpha)^{*} \phi_{s'}(\alpha)2\sin^2(2\alpha) d\alpha d\vec{\Omega}_{r}d\vec{\Omega}_{\rho}.
\end{eqnarray}

To calculate the Ramsey signal, Eq. (\ref{eq:RamseySignal}), we need the wavefunction overlaps, i.e. $\bra{F_{qs}(R)}\ket{F_{q's'}(R)}$ and $\bra{\phi_{s}(\alpha)}\ket{\phi_{s'}(\alpha)}$. Whether $s$ is imaginary or not does not change the functional form of the hyperangular wavefunction, $\phi_{s}(\alpha)$, unlike the hyperradial wavefunction, $F_{qs}(R)$. For the hyperangular integral there is only one case, but for the hyperradial integral there are three; the universal-universal, universal-Efimov and Efiomv-Efimov. We begin by considering the hyperangular integral.

The presence of the permutation operators makes evaluating the hyperangular integral directly difficult. To evaluate we transform the permuted terms into the same Jacobi set as the unpermuted term \cite{nielsen2001three}. However this limits us to the $l=0$ case, if the spherical harmonic term is non-constant then the coordinate transform is more complicated and the integral becomes intractable. The hyperangular integral is given \cite{fedorov1993efimov,fedorov2001regularization, braaten2006universality, thogersen2009universality}
\begin{eqnarray}
\bra{\phi_{0s}}\ket{\phi_{0s_{\rm i}}}&=&8\pi\int_{0}^{\pi/2}\left(\left(1+\hat{P}_{23}+\hat{P}_{13}\right)\frac{\varphi_{s}(\alpha)}{\sin(2\alpha)}\right)^{*}\left(\left(1+\hat{P}_{23}+\hat{P}_{13}\right)\frac{\varphi_{s_{\rm i}}(\alpha)}{\sin(2\alpha)}\right)\sin^2(2\alpha)d\alpha,\nonumber\\
&=&24\pi\Bigg[\int_{0}^{\pi/2}\varphi_{s}^{*}(\alpha)\varphi_{s_{\rm i}}(\alpha)d\alpha+\frac{4}{\sqrt{3}}\int_{0}^{\pi/2}\varphi_{s}^{*}(\alpha)\left[ \int_{|\pi/3-\alpha|}^{\pi/2-|\pi/6-\alpha|} \varphi_{s_{\rm i}}(\alpha') d\alpha' \right]  d\alpha \Bigg].\label{eq:HypAngInt}
\end{eqnarray}

For $l=0$ we have \cite{werner2008trapped, fedorov2001regularization}
\begin{eqnarray}
\varphi_{0s}\propto\sin\left(s\left(\frac{\pi}{2}-\alpha\right)\right).
\label{eq:AngWavefunc}
\end{eqnarray}
Note that evaluating Eq. (\ref{eq:HypAngInt}) with Eq. (\ref{eq:AngWavefunc}) does not give the same result as Ref. \cite{werner2008trapped} in general. This is because Ref. \cite{werner2008trapped} is firstly concerned with the overlaps with the ground state and secondly combine Eq. (\ref{eq:HypAngInt}) with Eq. (\ref{eq:Transcendental}) and so the results presented here and in the latter reference agree when $s$ is a unitary eigenvalue and $(q_{\rm i},s_{\rm i})=(0,2)$. Note that different hyperangular states of the same regime (i.e. two different unitary values of $s$ or two different non-interacting values of $s$) are orthogonal, but there is non-zero overlap between unitary and non-interacting states.

The hyperradial integrals are integrals of products of well understood functions. The universal-universal integral is given \cite{srivastava2003remarks}
\begin{eqnarray}
&&\bra{F_{qs}(R)}\ket{F_{q_{\rm i}s_{\rm i}}(R)}=\nonumber\\
&&\frac{a_{\mu}^2}{2}\binom{q+s}{q}\binom{q_{\rm i}+\dfrac{s_{\rm i}-s}{2}-1}{q_{\rm i}}\Gamma\left(\frac{s+s_{\rm i}}{2}+1\right){}_{3}F_{2}\bigg(-q,\frac{s+s_{\rm i}}{2}+1,\frac{s-s_{\rm i}}{2}+1;s+1,\frac{s-s_{\rm i}}{2}-q_{\rm i}+1;1\bigg),
\end{eqnarray}
the Efimov-Efimov \cite{gradshteyn2014table}
\begin{eqnarray}
\bra{F_{qs}(R)}\ket{F_{q_{\rm i}s}(R)} &=& a_{\mu}^2 \Re \Bigg[ \frac{\Gamma\left(s+1\right)\Gamma(-s)}{\Gamma\left(\dfrac{1-E_{q_{\rm i}}/\hbar\omega-s}{2}\right)\Gamma\left(\dfrac{3-E_{q}/\hbar\omega+s}{2}\right)}\nonumber\\
&&\times{}_{3}F_{2}\bigg(s+1,1,\frac{1-E_{q_{\rm i}}/\hbar\omega+s}{2}; 1+s,\frac{3-E_{q}/\hbar\omega+s}{2};1\bigg) \Bigg],
\end{eqnarray}
and the universal-Efimov \cite{gradshteyn2014table}
\begin{eqnarray}
&&\bra{F_{qs}(R)}\ket{F_{q_{\rm i}s_{\rm i}}(R)}=\frac{a_{\mu}^2}{4}\frac{(-1)^{q_{\rm i}}}{\Gamma(1+q_{\rm i})}\Bigg[\nonumber\\
&&\frac{\Gamma\left(\dfrac{2-s^{*}+s_{\rm i}}{2}\right)\Gamma\left(\dfrac{2+s^{*}+s_{\rm i}}{2}\right)\Gamma(-s_{\rm i})}{\Gamma(-q_{\rm i}-s_{\rm i})\Gamma\left(\dfrac{3-E_{q}/\hbar\omega+s_{\rm i}}{2}\right)}{}_{3}F_{2}\bigg(1+\frac{s_{\rm i}-s^{*}}{2},1+\frac{s^{*}+s_{\rm i}}{2},-q_{\rm i};1+s_{\rm i},\frac{3-E_{q}/\hbar\omega+s_{\rm i}}{2};1\bigg)\nonumber\\
&&+\frac{\Gamma\left(\dfrac{2+s^{*}-s_{\rm i}}{2}\right)\Gamma\left(\dfrac{2-s^{*}-s_{\rm i}}{2}\right)\Gamma(s_{\rm i})}{\Gamma(-q_{\rm i})\Gamma\left(\dfrac{3-E_{q}/\hbar\omega-s_{\rm i}}{2}\right)}{}_{3}F_{2}\bigg(1+\frac{s^{*}-s_{\rm i}}{2},1-\frac{s^{*}+s_{\rm i}}{2},-q_{\rm i}-s_{\rm i};1-s_{\rm i},\frac{3-E_{q}/\hbar\omega-s_{\rm i}}{2};1\bigg)\Bigg],
\end{eqnarray}
where we have used the identity
\begin{eqnarray}
L_{n}^{\alpha}(z)=\frac{(-1)^n}{n!}e^{z/2}z^{-(n+1)/2}W_{\dfrac{2n+\alpha+1}{2},\dfrac{\alpha}{2}}(z).%
\end{eqnarray}
For the hyperradial integral we have that for $s=s'$ the integral is 0 for $q\neq q'$.

To calculate the particle separation expectation value, Eq. (\ref{eq:ExpectR}), we again need to calculate a number of integrals involving the wavefunction. All the needed integrals except $\bra{F_{q's}}\tilde{R}\ket{F_{qs}}$ are given above. Previously for the hyperradial integral we had three cases, here we do not need to consider the universal-Efimov case as $s$ is the same in both bra and ket due to the orthogonality in $s$ of the hyperangular integral. For the universal-universal case we have \cite{srivastava2003remarks}
\begin{eqnarray}
\bra{F_{qs}(R)}\tilde{R}\ket{F_{q's}(R)}=\frac{a_{\mu}^2}{2}\binom{q+s}{q}\binom{q'-\frac{3}{2}}{q'}\Gamma\left(s+\frac{3}{2}\right){}_{3}F_{2}\left(-q,s+\frac{3}{2},\frac{3}{2};s+1,\frac{3}{2}-q';1\right),
\end{eqnarray}
and for the Efimov-Efimov case we have \cite{gradshteyn2014table}
\begin{eqnarray}
\bra{F_{qs}(R)}\tilde{R}\ket{F_{q's}(R)} &=& a_{\mu}^2 \Re \Bigg[ \frac{\Gamma\left(\dfrac{3}{2}+s\right)\Gamma\left(\dfrac{3}{2}\right)\Gamma(-s)}{\Gamma\left(\dfrac{1-E_{q'}/\hbar\omega -s}{2}\right)\Gamma\left(\dfrac{4-E_{q}/\hbar\omega +s}{2}\right)}  \nonumber\\
&&\times{}_{3}F_{2}\bigg(\frac{3}{2}+s,\frac{3}{2},\frac{1-E_{q'}/\hbar\omega +s}{2};1+s,\frac{4-E_{q}/\hbar\omega +s}{2};1\bigg) \Bigg].
\end{eqnarray}

\twocolumngrid
\bibliographystyle{apsrev4-1}{}
\bibliography{Few-Body-Refs}

\end{document}